\documentclass[useAMS,usenatbib,usegraphicx]{mn2e}
\usepackage{times}

\usepackage{graphicx}

\newcommand{{\qc}}{q_{\rm c}}
\newcommand{{\Rin}}{R_{\rm in}}
\newcommand{{\Mb}}{M_{\rm b}}
\newcommand{{\Mp}}{M_{\rm p}}
\newcommand{{\Ms}}{M_{\rm s}}

\title[]{Evolution of binary seeds in collapsing protostellar gas clouds}
\author[Satsuka, Tsuribe, Tanaka, and Nagamine]
{Tatsuya~Satsuka,$^{1,2}$\thanks{E-mail:satsuka@vega.ess.sci.osaka-u.ac.jp}
Toru~Tsuribe,$^{2}$ Suguru~Tanaka,$^1$ and Kentaro~Nagamine$^{1,3}$\\
$^{1}$Theoretical Astrophysics, Department of Earth and Space Science, Graduate School of Science, Osaka University, 1-1 Machikaneyama, Toyonaka 560-0043, Japan\\
$^{2}$ College of Science, Ibaraki University, 2-1-1 Bunkyo, Mito, Ibaraki 310-8512, Japan\\
$^{3}$ Department of Physics and Astronomy, University of Nevada, Las Vegas, 4505 S. Maryland Pkwy, Las Vegas, NV 89154-4002, USA
}
\begin{document}
\bibliographystyle{mn2e}
\date{Accepted xxx. Received xxx; in original form xxx}

\pagerange{\pageref{firstpage}--\pageref{lastpage}} \pubyear{xxxx}

\maketitle

\label{firstpage}

\begin{abstract}
We perform three dimensional smoothed particle hydrodynamics (SPH) 
simulations of gas accretion onto the seeds of binary stars to investigate their short-term evolution.
Taking into account of dynamically evolving envelope with non-uniform distribution of gas density and angular momentum of accreting flow,
our initial condition includes a seed binary and a surrounding gas envelope, modelling the phase of core collapse of gas cloud when the fragmentation has already occurred.
We run multiple simulations with different values of initial mass ratio $q_0$ (the ratio of secondary over primary mass) and gas temperature.  For our simulation setup, we find a critical value of $\qc=0.25$ which distinguishes the later evolution of mass ratio q as a function of time.
If $q_0 \ga \qc$, the secondary seed grows faster and $q$ increases monotonically towards unity.
If $q_0 \la \qc$, on the other hand, the primary seed grows faster and $q$ is lower than $q_0$ at the end of the simulation. 
Based on our numerical results, we analytically calculate the long-term evolution of the seed binary including the growth of binary by gas accretion.
We find that the seed binary with $q_0 \ga \qc$ evolves towards an equal-mass binary star, and that with $q_0 \la \qc$ evolves to a binary with an extreme value of $q$.
Binary separation is a monotonically increasing function of time for any $q_0$, suggesting that the binary growth by accretion 
does not lead to the formation of close binaries. 
\end{abstract}


\section[]{Introduction}
It is widely recognised that the majority of main sequence stars and pre-main sequence stars are in binary systems and that these binaries have various distributions in mass ratio $q$, binary separation, and binary frequency \citep{Duquennoy_Mayor_91, Ghez_etal_93, Kouwenhoven_etal_05,Raghavan_etal_10, Kraus_etal_11, DeRosa_etal_14}.
For main sequence stars, \cite{Duquennoy_Mayor_91} provided statistics of G-dwarf binary systems.
They found that binaries have log-normal separation distribution, and the number of binaries with wide separations decreases with increasing $q$, while the number of binaries with close separations is roughly constant or increase with increasing $q$.
\cite{Raghavan_etal_10} and \cite{DeRosa_etal_14} found a similar distribution of $q$ for A-type binaries as \cite{Duquennoy_Mayor_91}. 

For young stellar objects (YSOs), the number of binaries with intermediate mass YSOs decreases with increasing $q$ \citep{Kouwenhoven_etal_05}.
In contrast, the number of binaries with low-mass YSOs increases with increasing $q$ \citep{Kraus_etal_11}, and have a log-normal separation distribution.
Although observers have investigated various distributions of binary properties in detail as mentioned above, formation process of these binaries is not well understood.

The favoured scenario of binary formation is the fragmentation during runaway collapse of cloud core
 \citep{Boss_Bodenheimer_79,Miyama_etal_84,Tsuribe_Inutsuka_99b,Tsuribe_Inutsuka_99a},  
and the fragmentation in protostellar disc after the runaway collapse \citep{Williams_Tohline_88,Adams_etal_89,Bonnell_94,Bonnell_Bate_94b,Woodward_etal_94,Vorobyov_10}.
Details of the fragmentation process during cloud core collapse is described in \cite{Tohline_02} and the references therein.
In the present paper, we do not discuss the fragmentation process, but focus on the evolution of a seed binary after fragmentation.

After the fragmentation, the binary seeds start to accrete the surrounding envelope and grow towards main sequence stars.
Since the initial mass of seed binary is lower than $1$ per cent of the stellar mass \citep{Bonnell_Bate_94b}, observed physical properties of a binary (i.e., properties at the end of accretion phase) are completely different from those of the seed binaries. 

Recently, large-scale hydrodynamical simulations have been performed in order to explain the observed binary properties \citep{Bate_etal_02b,Bate_etal_02c,Bate_Bonnell_Bromm_03,Attwood_etal_09,Bate_09a,Bate_09b,Offner_etal_09}. 
\cite{Bate_09a} simulated a large-scale, homogeneous, isothermal, and turbulent cloud core collapse ignoring radiative and magnetic effects, and succeeded in reproducing the observed multiplicity and binary properties such as the frequency of very low-mass binaries, the log-normal separation distribution, and the $q$-distribution of wide and close binaries.
However, they did not mention about how and what physical processes determine the distributions of binaries.
In order to understand that, we need to investigate the gas accretion onto a binary in more detail.

A number of simulations of gas accretion onto a binary have been performed using SPH codes in two dimensions (2D) \citep{Dunhill_etal_15,Young_etal_15,Young_Clarke_15} and 3D \citep{Artymowicz_Lubow_96,Bate_Bonnell_97},  as well as using grid codes in 2D \citep{Ochi_etal_05,Hanawa_etal_10,D'Orazio_etal_13,Farris_etal_14}.
For example, \cite{Bate_Bonnell_97} investigated a steady, isothermal, non-self-gravitating gas accretion onto a binary assuming 
constant angular momentum and density of gas at the outer boundary.
They found that, in the case of low gas angular momentum, the $q$-value decreases, because the primary is closer to the mass centre of binary than the secondary. 
On the other hand, in the case of high angular momentum, the $q$-value increases because the infalling gas encounters the secondary first.

\cite{Ochi_etal_05} also investigated in a similar model to \cite{Bate_Bonnell_97}, 
and found contradictory results that the $q$-value can decrease even when the angular momentum of gas is high.
They argued that this discrepancy was caused by the lower numerical resolution in the simulation of \cite{Bate_Bonnell_97}, 
which may have enhanced the accretion onto the secondary.  
However, there were other differences between the simulations of \cite{Bate_Bonnell_97} and \cite{Ochi_etal_05}, 
such as the gas temperature, computing method, and gravitational potential of the binary, 
therefore the ultimate cause of the difference was unclear at that time. 

\cite{Young_etal_15} returned to this problem and investigated the dependence on gas temperature in the same model as \cite{Bate_Bonnell_97} and \cite{Ochi_etal_05}, 
where they fixed the angular momentum of gas and changed only the gas temperature.
As a result, they concluded that the disagreement between \cite{Bate_Bonnell_97} and \cite{Ochi_etal_05} was simply caused by the difference in the gas temperature.
\cite{Young_etal_15} found that, in the case of cold gas, the gas from circum-binary disc is easily trapped inside the secondary's Roche lobe, whereas in the hot case, the flow from the secondary's Roche lobe to the primary's Roche lobe emerges, suppressing the growth of $q$-value. 

These previous works that we described above were limited to the most simple situations, i.e., 
isothermal non-self-gravitating gas, constant angular momentum and density of gas at the outer boundary, 
and an isolated binary (i.e., no growth of binary by accretion).
However, if we want to compare the simulations with real binary systems in star-forming regions, we need to consider the unsteady gas accretion caused by the non-uniform distribution of angular momentum and density of the envelope.
Motivated by this current situation, we perform SPH simulations to investigate the unsteady gas accretion onto a seed binary considering 
the non-uniform distribution of angular momentum and density of infalling envelope. 
Note that we still ignore the self-gravity of gas and the growth of binary by accretion, with the aim of understanding physically how these distributions of gas affect the binary evolution. 
If the accreted mass exceeds the initial mass of the seed binary, the self-gravity of gas and the growth of binary by accretion might cause non-negligible effects on binary evolution. 
Therefore, we focus on the short-term evolution until the accreted mass exceeds the initial binary mass. 
Sections~\ref{sec:model} and \ref{sec:method} describe our model and calculation method. 
We present the results of our simulations in Section~\ref{sec:results}.  
In Section \ref{sec:discussion}, we discuss the results 
and give estimates of the long-term evolution of seed binaries. 
We conclude in Section~\ref{sec:conclusion}.

\section[]{MODEL OF SEED BINARY AND ENVELOPE}\label{sec:model}

\begin{figure}
    \begin{center}
      \includegraphics[width=0.8\columnwidth]{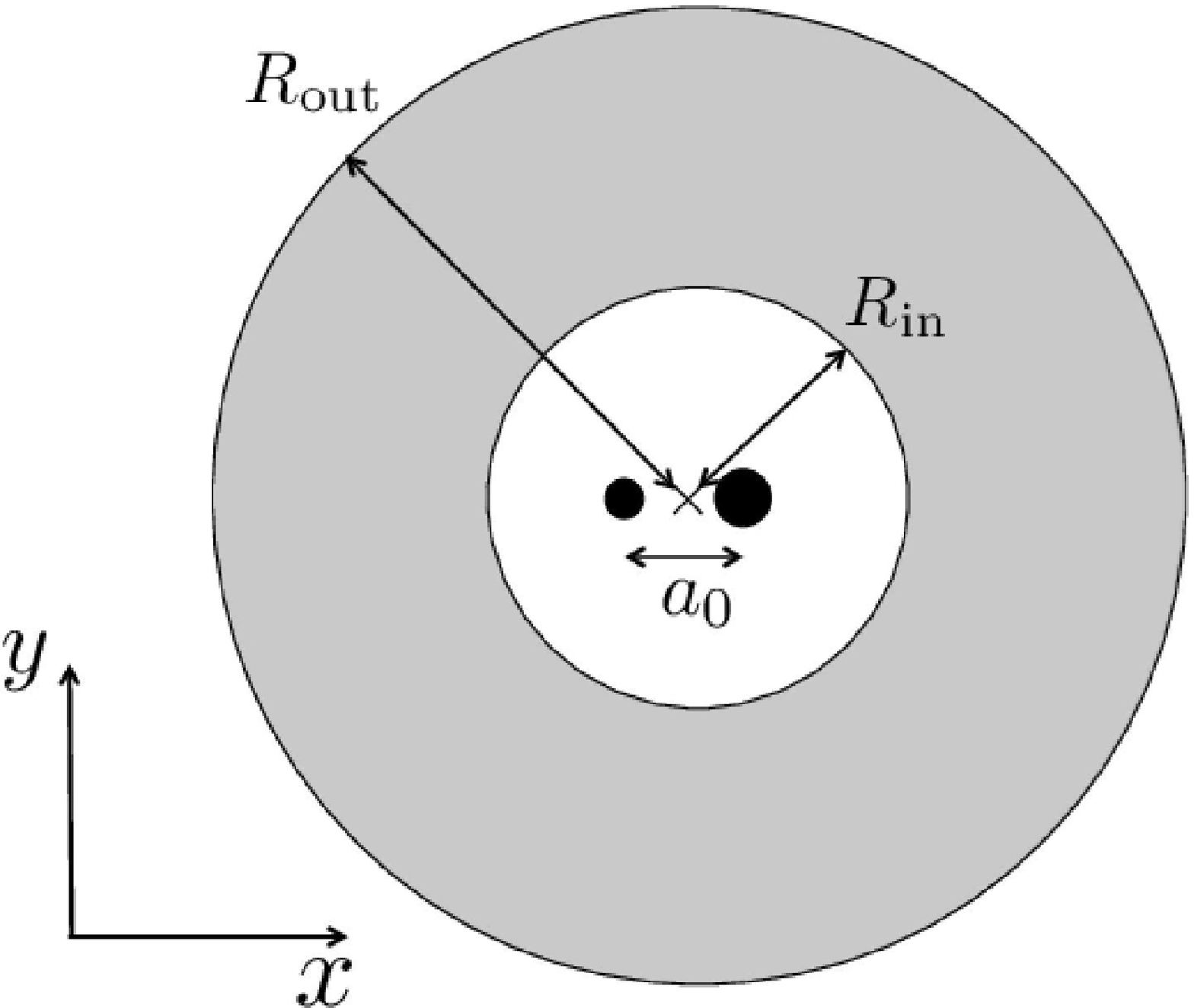}
    \end{center}
      \vspace{5mm}
    \begin{center}
      \includegraphics[width=0.8\columnwidth]{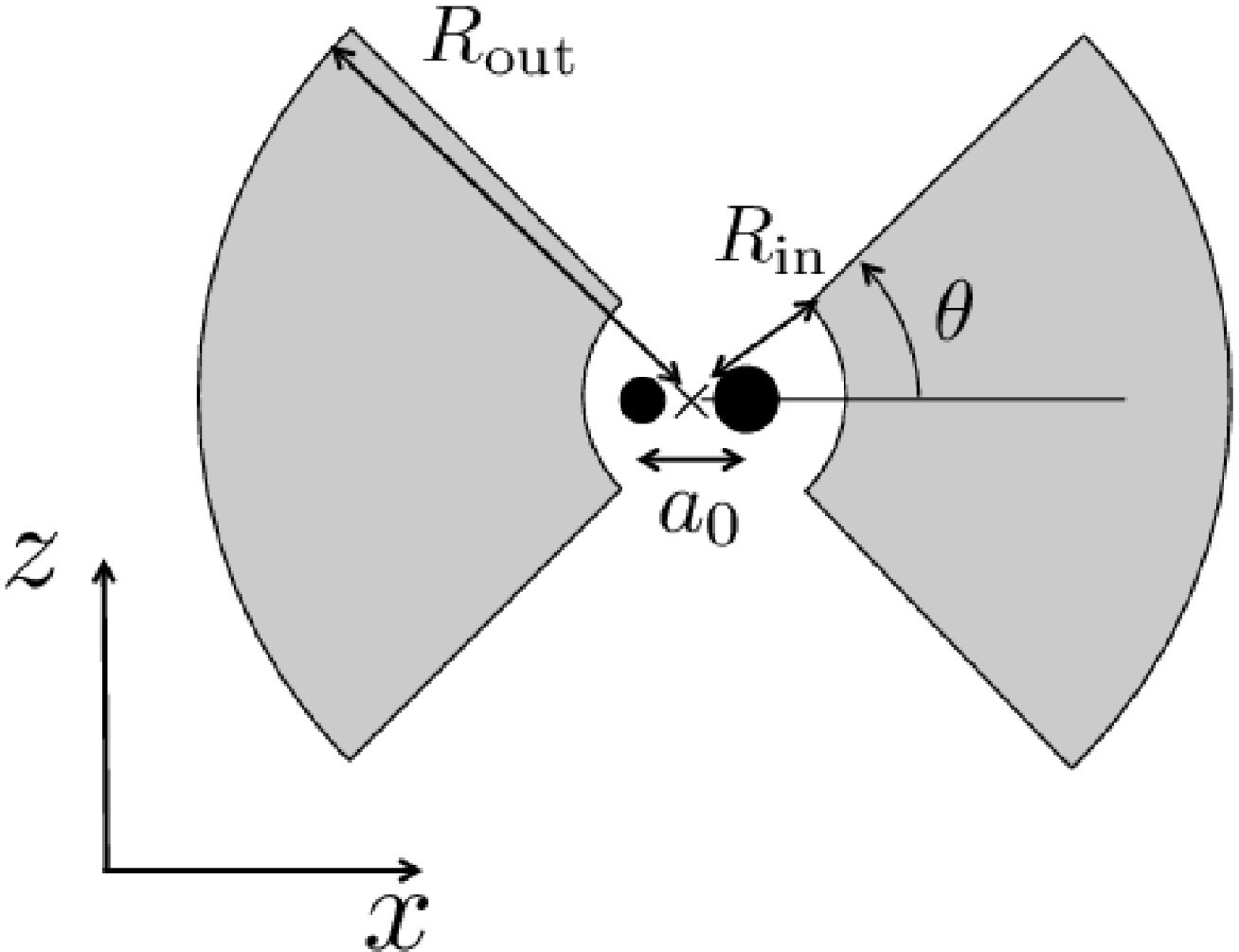}
    \end{center}
  \caption{Schematic diagrams of the initial condition of our simulations in the centre-of-mass frame of the envelope and binary. 
  The cross shows the origin,  two black circles in the centre are the binary seeds, and the gray regions show the envelope. The top (bottom) panel shows the cross section in a face-on (edge-on) view. The specific angular momentum vector of gas is aligned with $z$-axis.
The angle $\theta$ is determined by equation~(\ref{eq:theta}).} 
  \label{fig:cartoon}
\end{figure}

In the present work, we skip the formation process of the seed binary in order to focus on the gas accretion onto the seed binary. 
The seed binary and the surrounding envelope are set up in the initial condition as we describe below
and in Fig.~\ref{fig:cartoon}.

For the density and specific angular momentum distribution of the envelope, 
we take the following profiles: 
\begin{eqnarray}
\rho(r) &=& \rho_{\rm in} \left( \frac{\Rin}{r} \right)^2 \quad {\rm and} \label{eq:rho_des}\\
j(r) &=& j_{\rm in}\left(\frac{r}{\Rin} \right),
\label{eq:j_des}
\end{eqnarray}
where $r$ is distance from the centre-of-mass of the envelope, the constants $\rho_{\rm in}$ and $j_{\rm in}$ represent the density and specific angular momentum at $r = \Rin$, 
and $\Rin$ is the radius of the inner edge of the envelope. 
This power-law distribution is motivated by the numerical results often seen for the self-gravitating isothermal cloud core collapse with rotation \citep{Narita_etal_84,Matsumoto_etal_97,Saigo_Hanawa_98}.

Within the above gas distribution, we assume that the seed binary is formed via fragmentation of gas, 
carve out the envelope within $r< \Rin$, and convert its mass into the seed binary while conserving mass and angular momentum. 
The seed binary is modelled by two point masses, and their masses are 
$\Mp$ and $\Ms$ for the primary and secondary, respectively. 
The total mass of binary is $\Mb \equiv \Mp + \Ms$, 
the initial mass ratio is $q_0\equiv \Ms / \Mp$, and the initial binary separation is $a_0$.
We assume that these seeds move circularly around the centre-of-mass with Keplerian velocities. 
By conservation laws, the mass and specific angular momentum of the seed binary is represented by 
\begin{eqnarray}
\Mb &=& \int_{0}^{\Rin} 4\pi r^2 \rho(r) {\rm d}r = 4\pi {\Rin}^3 \rho_{\rm in}, \\
J_{\rm b} &=& \int_{0}^{\Rin} 4\pi r^2 \rho(r) j(r) {\rm d}r = \frac{1}{2} \Mb j_{\rm in}\nonumber\\
          &=& \frac{q_0}{(1+q_0)^2}j_{\rm circ} \Mb, \label{eq:J_b}
\end{eqnarray}
where $j_{\rm circ} \equiv \sqrt{G \Mb a_0}$ is the reference specific angular momentum.
Once $R_{\rm in}$ and $q_0$ are determined, $j_{\rm in}$ and $\rho_{\rm in}$ are given by 
\begin{eqnarray}
\rho_{\rm in} &=& \frac{\Mb}{4\pi {\Rin}^3},\\
j_{\rm in} &=& \frac{2q_0}{(1+q_0)^2}\left( G \Mb a_0 \right)^{1/2} \label{eq:j_0}.
\end{eqnarray}

The envelope gas at $r> \Rin$ is distributed as equations~(\ref{eq:rho_des}) and (\ref{eq:j_des}).
We set the velocities of gas at $\Rin < r < R_{\rm out}$ as 
\begin{equation}
\frac{j(r)^2}{r_{\rm cyl}^3} < \frac{G \Mb}{r} \frac{r_{\rm cyl}}{r},
\label{eq:theta}
\end{equation}
such that the gravitational force exceeds the centrifugal force (see Fig.~\ref{fig:cartoon}).
The initial radial velocity of gas is assumed as
\begin{equation} 
v_r(r) = \left( \frac{2G \Mb}{r} - \frac{{j(r)}^2}{{r_{\rm cyl}}^2}  \right)^{1/2},
\end{equation}
such that the kinetic energy of gas is equal to the gravitational energy, where $r_{\rm cyl}$ is cylindrical radius in the centre-of-mass frame.
For simplicity, we assume a non-self-gravitating isothermal gas, and no magnetic fields and radiation. 
We emphasize that the important difference between our work and the previous ones \citep{Bate_Bonnell_97,Ochi_etal_05,Young_etal_15,Young_Clarke_15} 
is that we aim to investigate unsteady evolution of binary by considering non-uniform distribution of gas density and angular momentum, while we still assume isolated binary and ignore self-gravity of gas.

\section[]{NUMERICAL METHOD}\label{sec:method}

We use {\tt GADGET-3} SPH code  (originally described by \citealt{Springel_05}) 
in three dimension. 
In this code, smoothing length of each gas particle is determined by the number of neighbour particles: $(4\pi/3)h^3\rho=N_{\rm ngb}m$, where $h$ is the smoothing length, $m$ is the mass of SPH particle, and $N_{\rm ngb}$ is the number of neighbour particles.
We adopt $N_{\rm ngb}=50$, which corresponds to $h\sim 2.3(m/\rho)^{1/3}$.
For calculation of hydrodynamics, we choose a polytropic index $\gamma=1$ assuming an isothermal gas, and adopt Monaghan-Balsara form of artificial viscosity with the parameters $\alpha = 1.0$ and $\beta=2.0$ (see \citealt{Springel_05} and references therein).
All of our simulations are performed with $128^3$ SPH particles, which is 
roughly two orders of magnitude higher than that in \cite{Bate_Bonnell_97}.
In some of our simulations, a steady circum-secondary disc forms, 
and the smoothing length is about one-tenth to one-fifth of the scale height at the outer edge of the disc, satisfying the criterion of \cite{Young_etal_15} (see Appendix).

The seed binary is treated as two sink particles with a sink radius of $R_{\rm sink}=0.01a_0$, 
following \citet{Young_etal_15}.
The SPH particles are removed from the computational domain once they 
fall into the sink radius of each seed or reach the outer boundary, without any feedback to the seeds. 

In the case where the envelope has an angular momentum, circum-stellar discs are formed \citep{Artymowicz_Lubow_96,Bate_97,Bate_Bonnell_97,Ochi_etal_05}.
We define our circum-stellar discs as the gas with $J < U_{\rm L1}$, where $J$ is the Jacobi constant of gas and $U_{\rm L1}$ is the sum of gravitational and centrifugal potential at L1 point in the corotating frame of the seed binary.
The Jacobi constant of gas is written by
\begin{eqnarray}
J = \frac{1}{2}v^2 - \frac{GM_{\rm p}}{r_{\rm p}} - \frac{GM_{\rm s}}{r_{\rm s}} + c_s^2 \,{\rm ln}\,\rho,
\end{eqnarray}
where $r_{\rm p}$, $r_{\rm s}$, $c_s$ are the distance from the primary, the distance from the secondary, and the sound speed of gas, respectively.

By our definition of the circum-stellar discs, it is possible that the gas in the circum-stellar discs will be accreted onto each seed via viscous evolution. 
Therefore we define the time-dependent mass ratio as 
\begin{equation}
q(t) = \frac{\Ms + \Delta\Ms(t) }{\Mp + \Delta\Mp(t)},
\end{equation}
where
\begin{eqnarray}
\Delta \Mp(t) &=& M_{\rm acc,p}(t) + M_{\rm disc,p}(t),\label{eq:delta_Mp}\\
\Delta \Ms(t) &=& M_{\rm acc,s}(t) + M_{\rm disc,s}(t). \label{eq:delta_Ms}
\end{eqnarray}
Here, $M_{\rm acc,p}$ is the mass accreted onto the primary, $M_{\rm acc,s}$ is the mass accreted onto the secondary, $M_{\rm disc,p}$ is the mass of circum-primary disc, and $M_{\rm disc,s}$ is the mass of circum-secondary disc. 
We terminate our simulations when $\Delta \Mb(t)\equiv \Delta \Mp(t) + \Delta \Ms(t)$ exceeds $\Mb$, because the self-gravity of gas and the growth of binary cannot be ignored after $\Delta \Mb(t)> \Mb$.

With these setup, we compute the gas accretion onto the seed binary for several values of $q_0$ in the range of $0.1<q_0<1.0$ and the sound speed $c_s/\sqrt{GM_{\rm b}/a_0} = 0.05$ (cold) and $0.25$ (hot). 
Hereafter, we use the units of $G=M_b=a_0=1$, 
in which the orbital period of the seed binary corresponds to $2\pi$. 


\section[]{RESULTS}\label{sec:results}
\subsection{Formation of discs}\label{subsec:results_disc}

\begin{figure*}
  \begin{minipage}{\columnwidth}
    \begin{center}
      \includegraphics[width=\columnwidth]{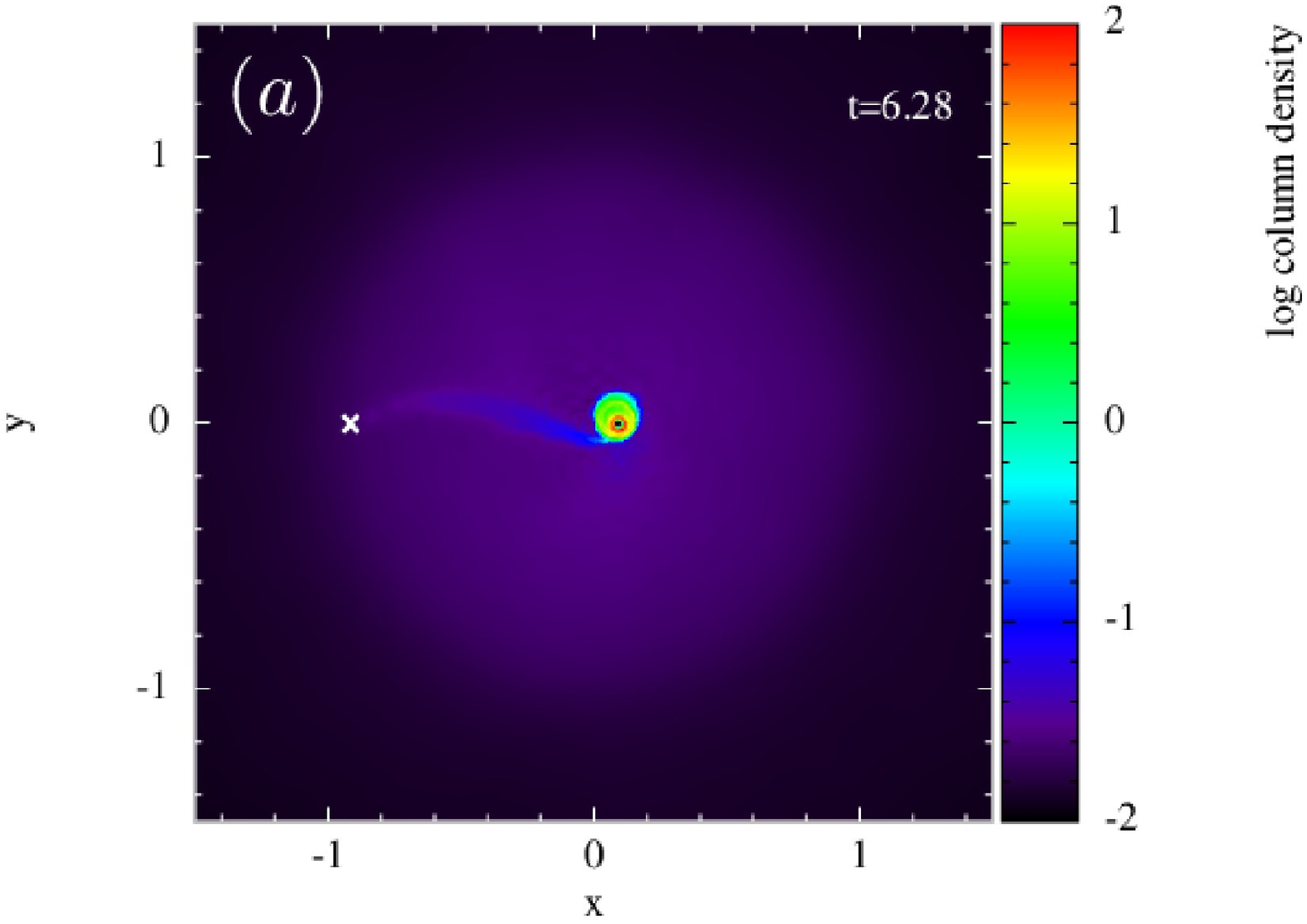}
    \end{center}
  \end{minipage}
  \begin{minipage}{\columnwidth}
    \begin{center}
      \includegraphics[width=\columnwidth]{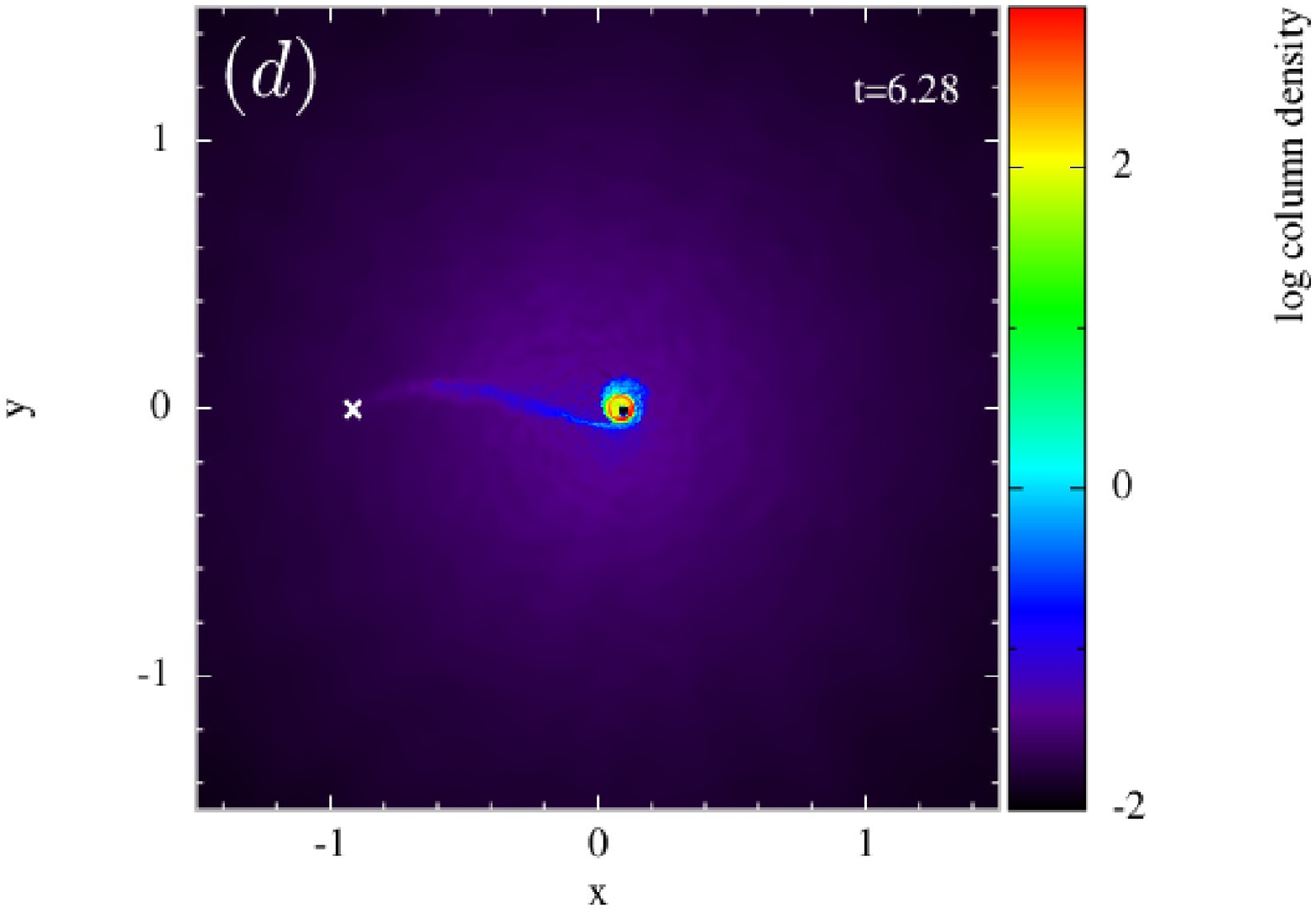}
    \end{center}
  \end{minipage}
  \begin{minipage}{\columnwidth}
    \begin{center}
      \includegraphics[width=\columnwidth]{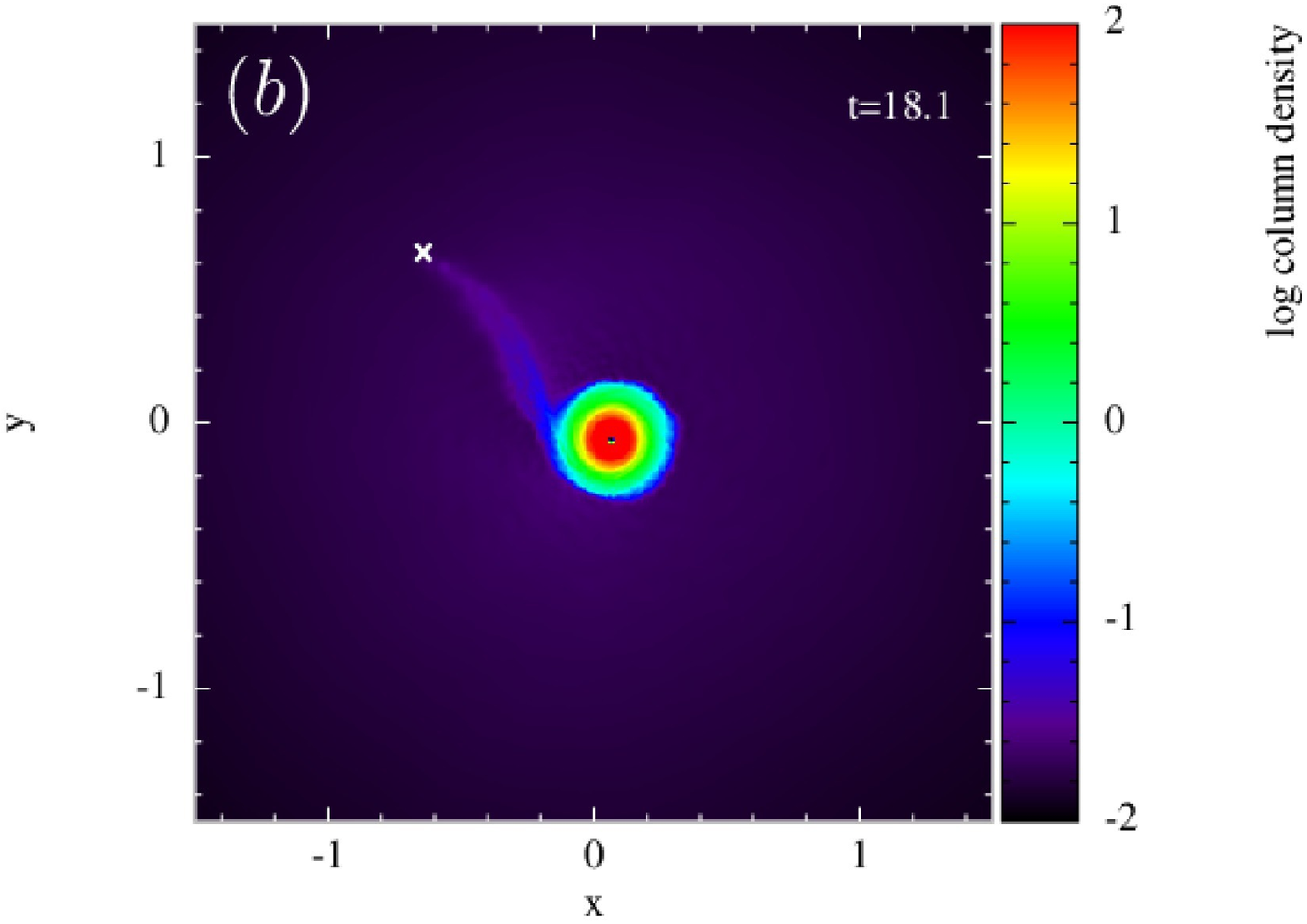}
    \end{center}
  \end{minipage}
  \begin{minipage}{\columnwidth}
    \begin{center}
      \includegraphics[width=\columnwidth]{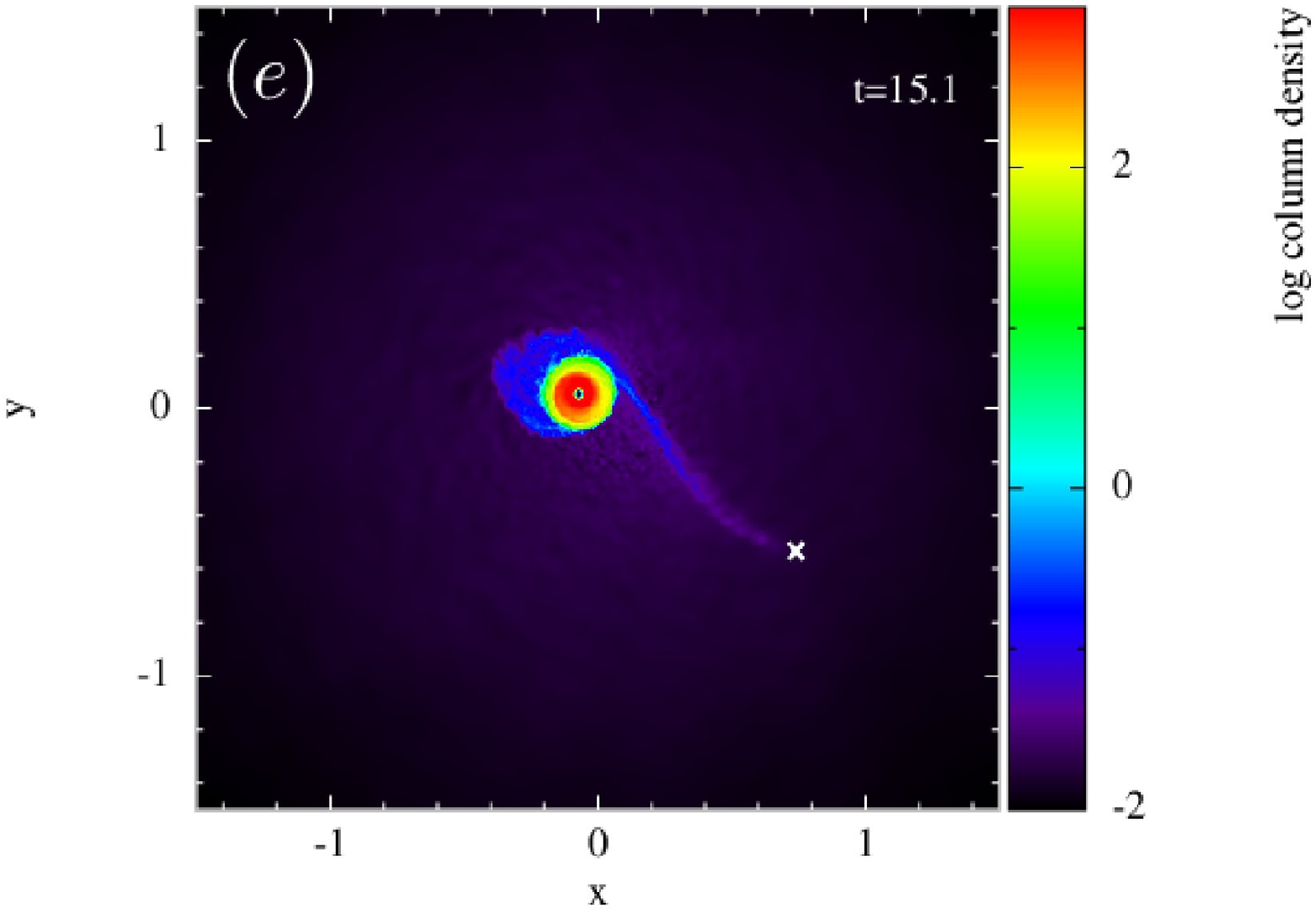}
    \end{center}
  \end{minipage}
  \begin{minipage}{\columnwidth}
  \vspace{-3mm}
    \begin{center}
      \includegraphics[width=\columnwidth]{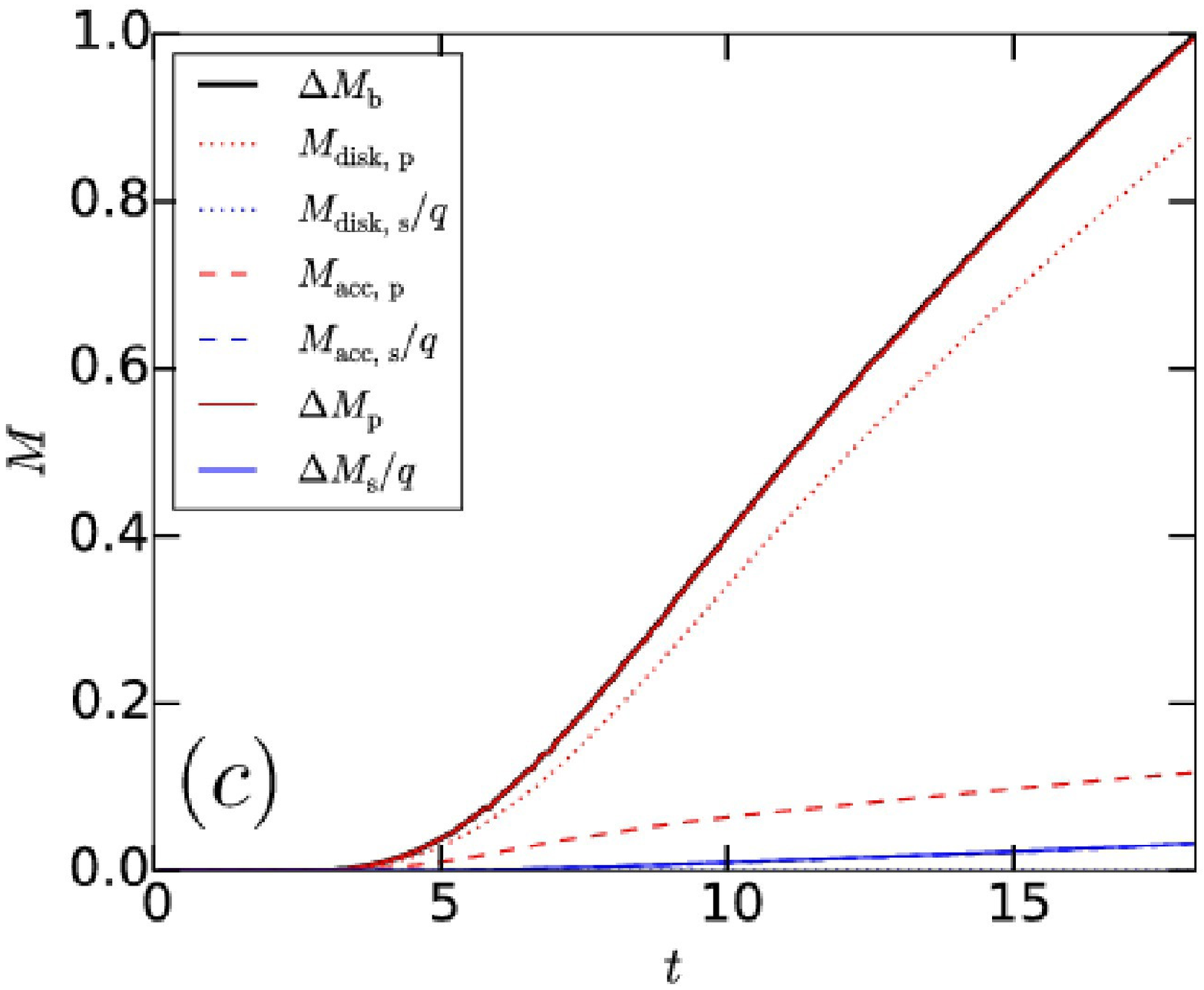}
    \end{center}
  \end{minipage}
  \begin{minipage}{\columnwidth}
  \vspace{-3mm}
    \begin{center}
      \includegraphics[width=\columnwidth]{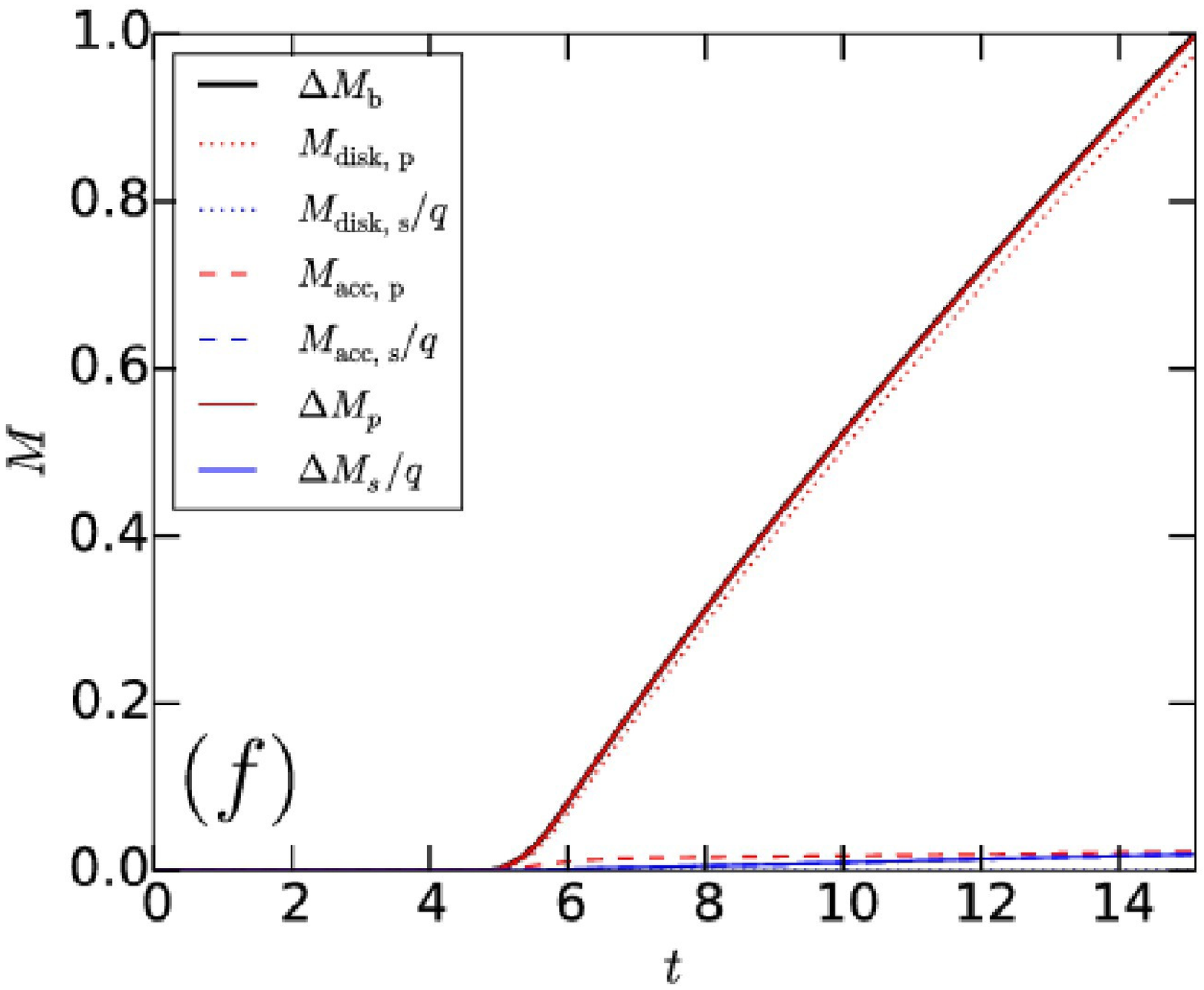}
    \end{center}
  \end{minipage}
  \caption{Time evolution of gas around the seed binary. The top and middle panels show the face-on logarithmic surface density maps in the centre-of-mass frame for $q_0=0.1$. The hot case is shown in the left column and the cold case on the right column.  The top panels are at $t=2\pi$, and the middle panels show the final snapshot at $t=18.1$ (panel ($b$)) and $t=15.1$ (panel ($e$)).
The white crosses show the positions of the secondary.
The bottom panels show the time evolution of mass accreted onto the primary $M_{\rm acc,p}$ (red dashed), mass accreted onto the secondary $M_{\rm acc,s}$ (blue dashed), mass of the circum-primary disc $M_{\rm disc,p}$ (red dotted), mass of the circum-secondary disc $M_{\rm disc,s}$ (blue dotted), change from the initial mass of primary $\Delta \Mp$ (red solid), change from the initial mass of secondary $\Delta \Ms$ (blue solid), change from the initial binary mass $\Delta \Mb$ (black solid).  
Note that all values for secondary is divided by the initial mass ratio $q_0$. 
All density maps in this paper are produced using {\tt SPLASH} visualization code \citep{Price_07}
}
  \label{fig:q01}
\end{figure*}

\begin{figure*}
  \begin{minipage}{\columnwidth}
    \begin{center}
      \includegraphics[width=\columnwidth]{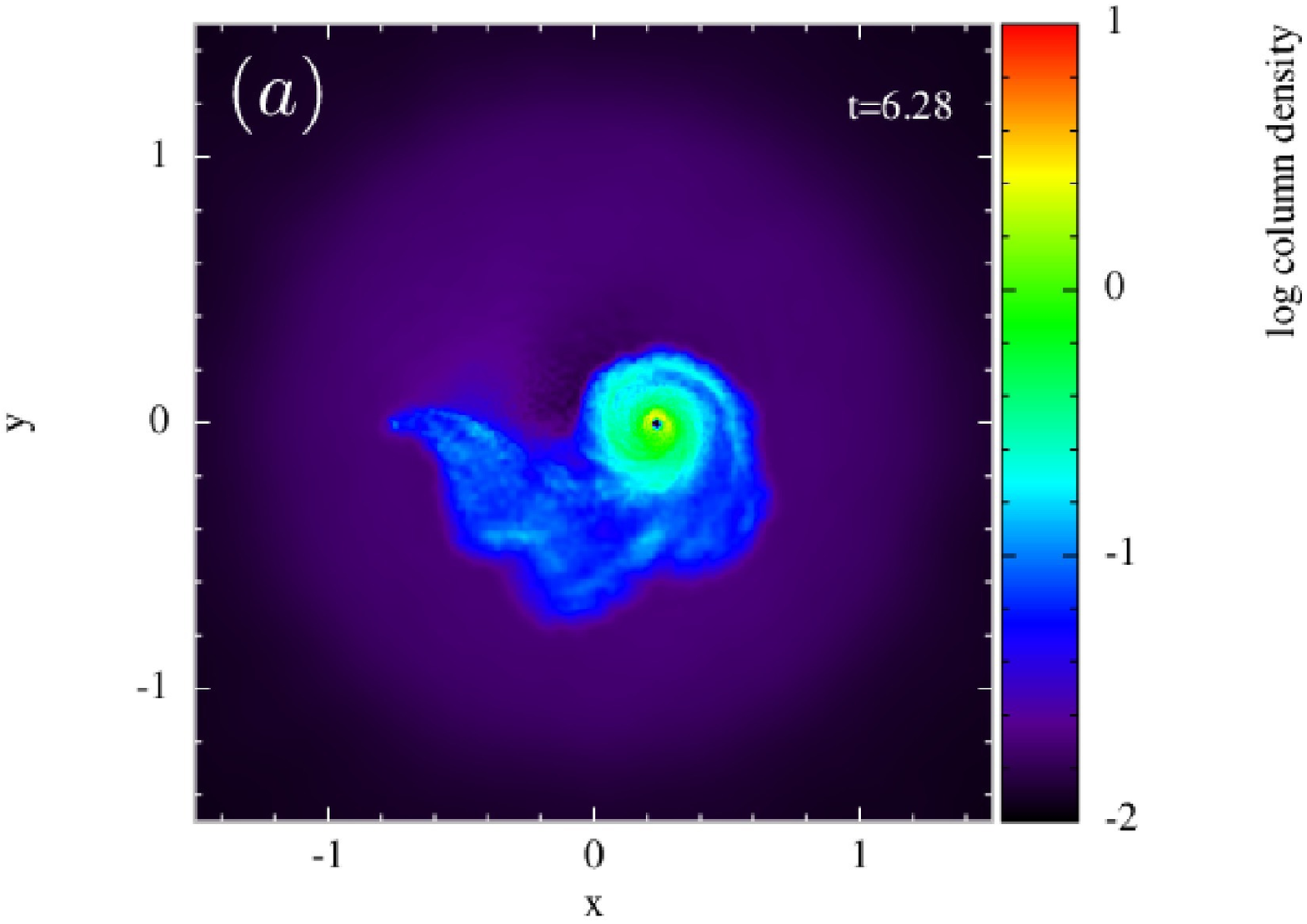}
    \end{center}
  \end{minipage}
  \begin{minipage}{\columnwidth}
    \begin{center}
      \includegraphics[width=\columnwidth]{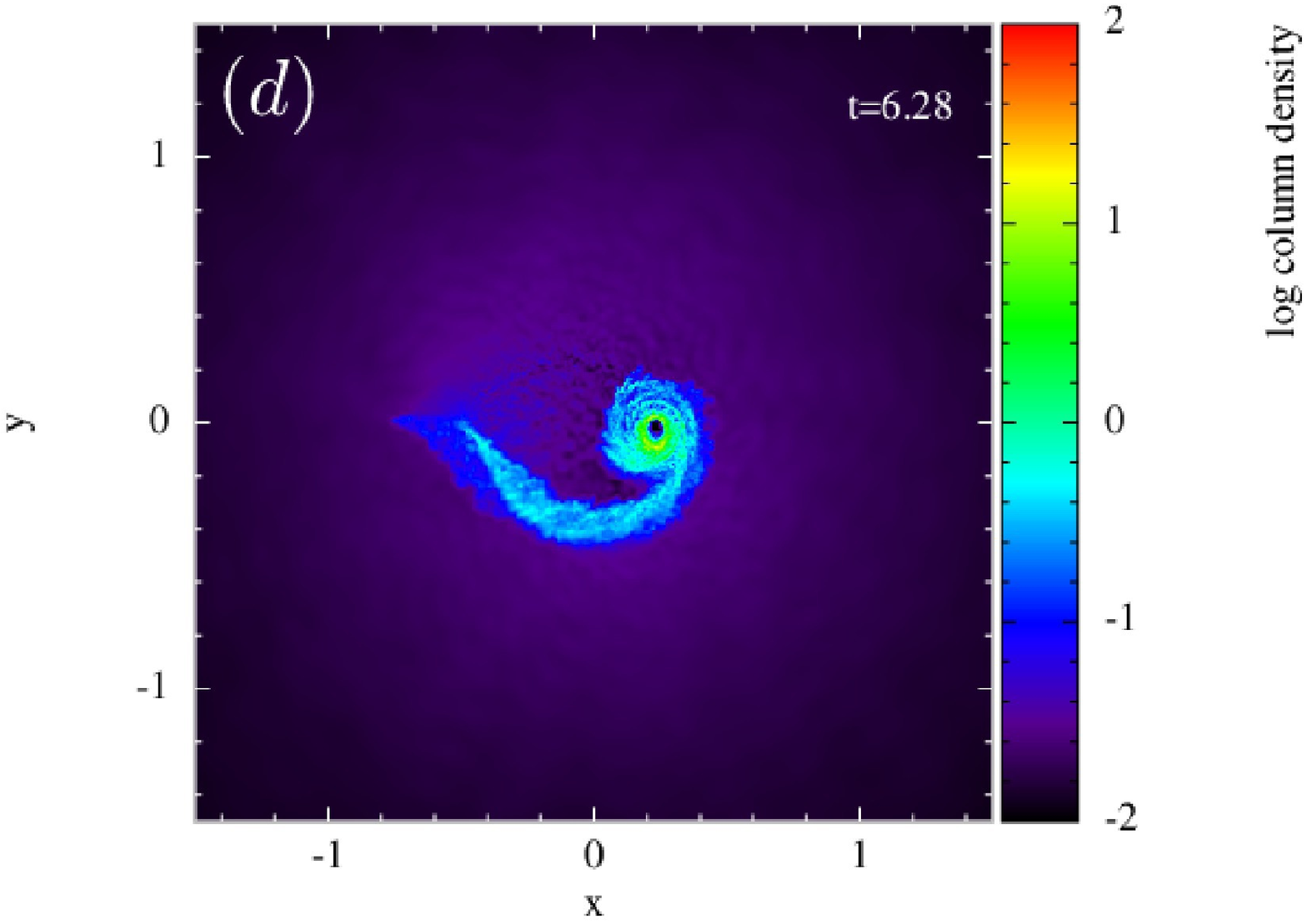}
    \end{center}
  \end{minipage}
  \begin{minipage}{\columnwidth}
    \begin{center}
      \includegraphics[width=\columnwidth]{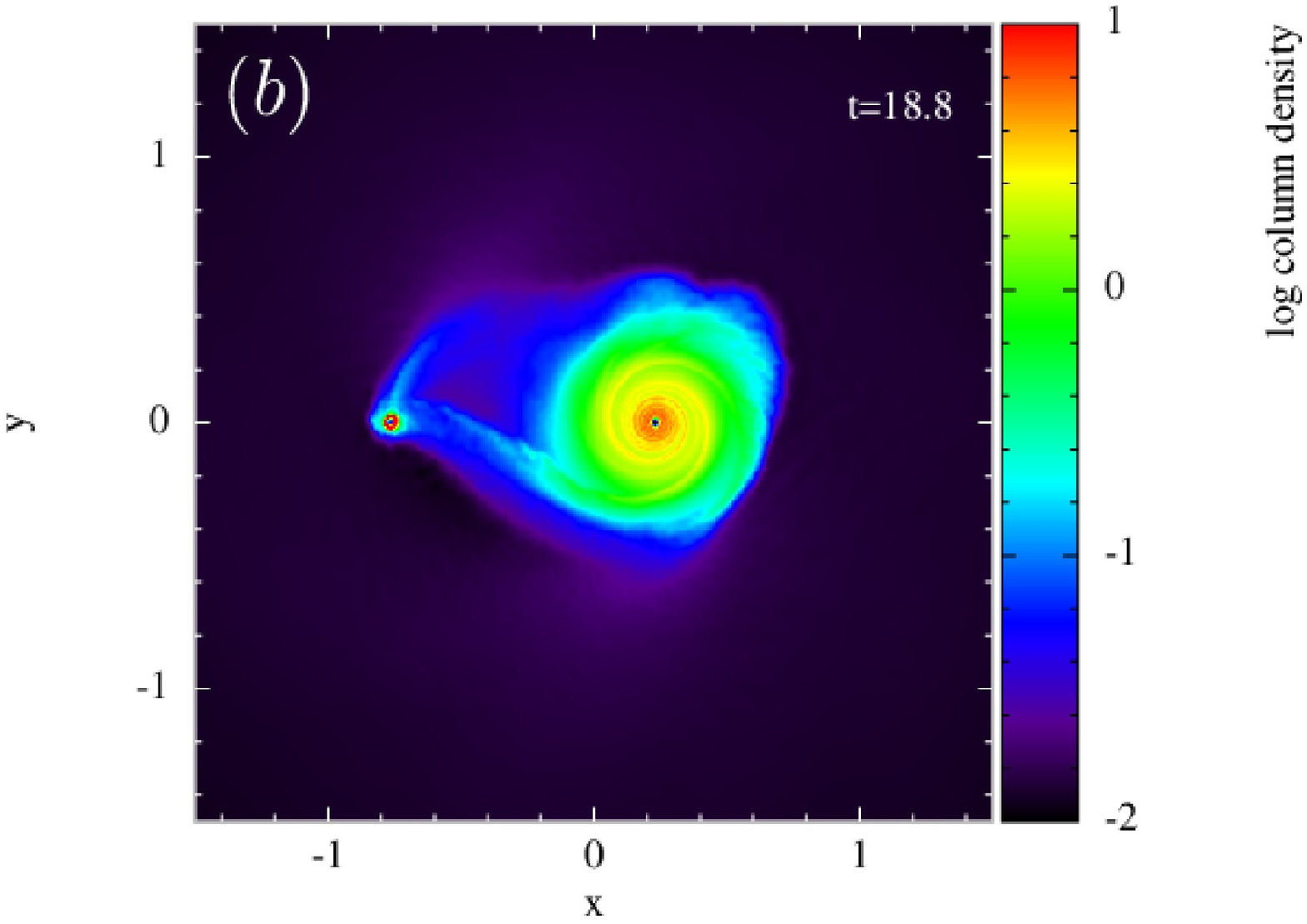}
    \end{center}
  \end{minipage}
  \begin{minipage}{\columnwidth}
    \begin{center}
      \includegraphics[width=\columnwidth]{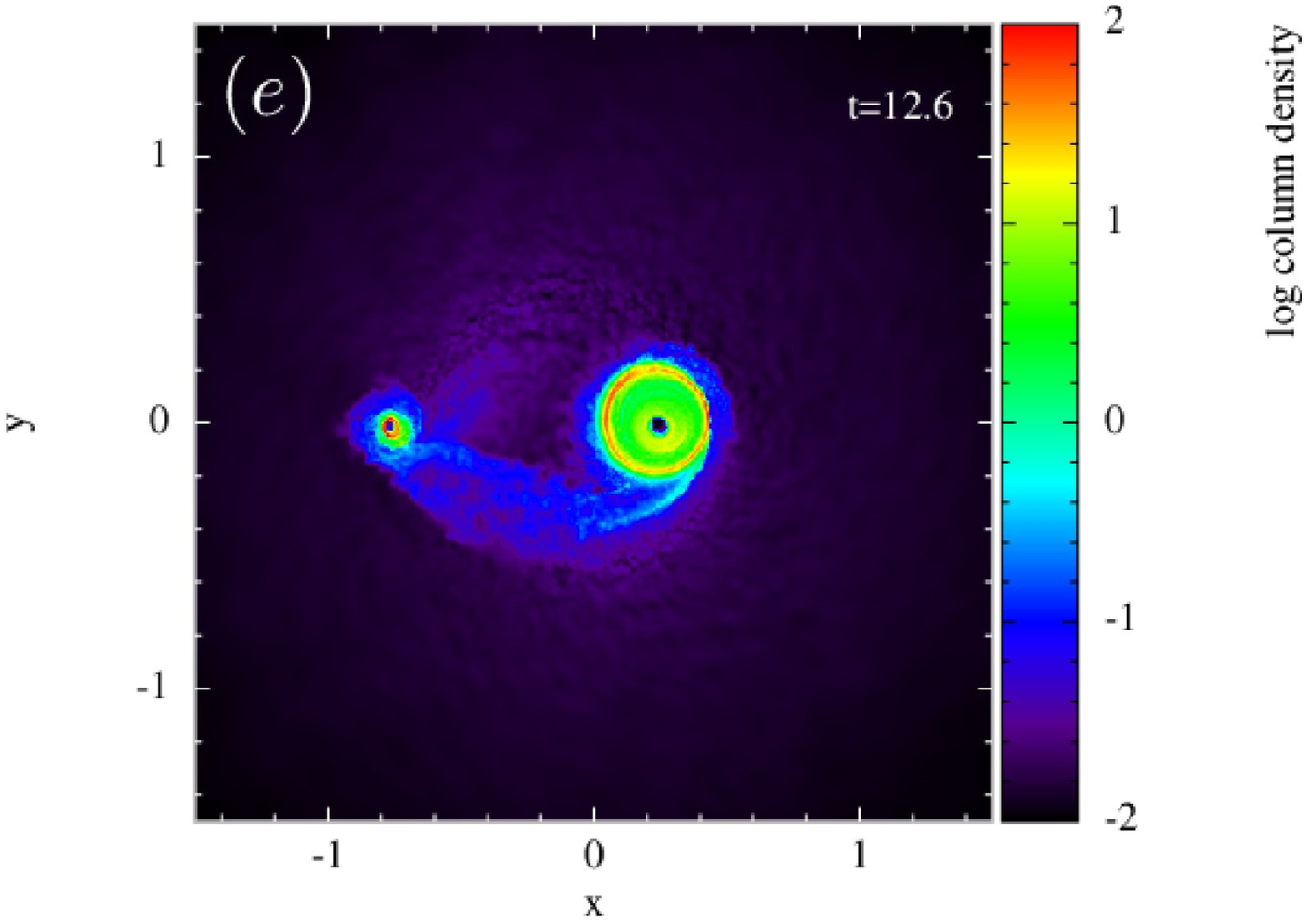}
    \end{center}
  \end{minipage}
  \begin{minipage}{\columnwidth}
    \vspace{-3mm}
    \begin{center}
      \includegraphics[width=\columnwidth]{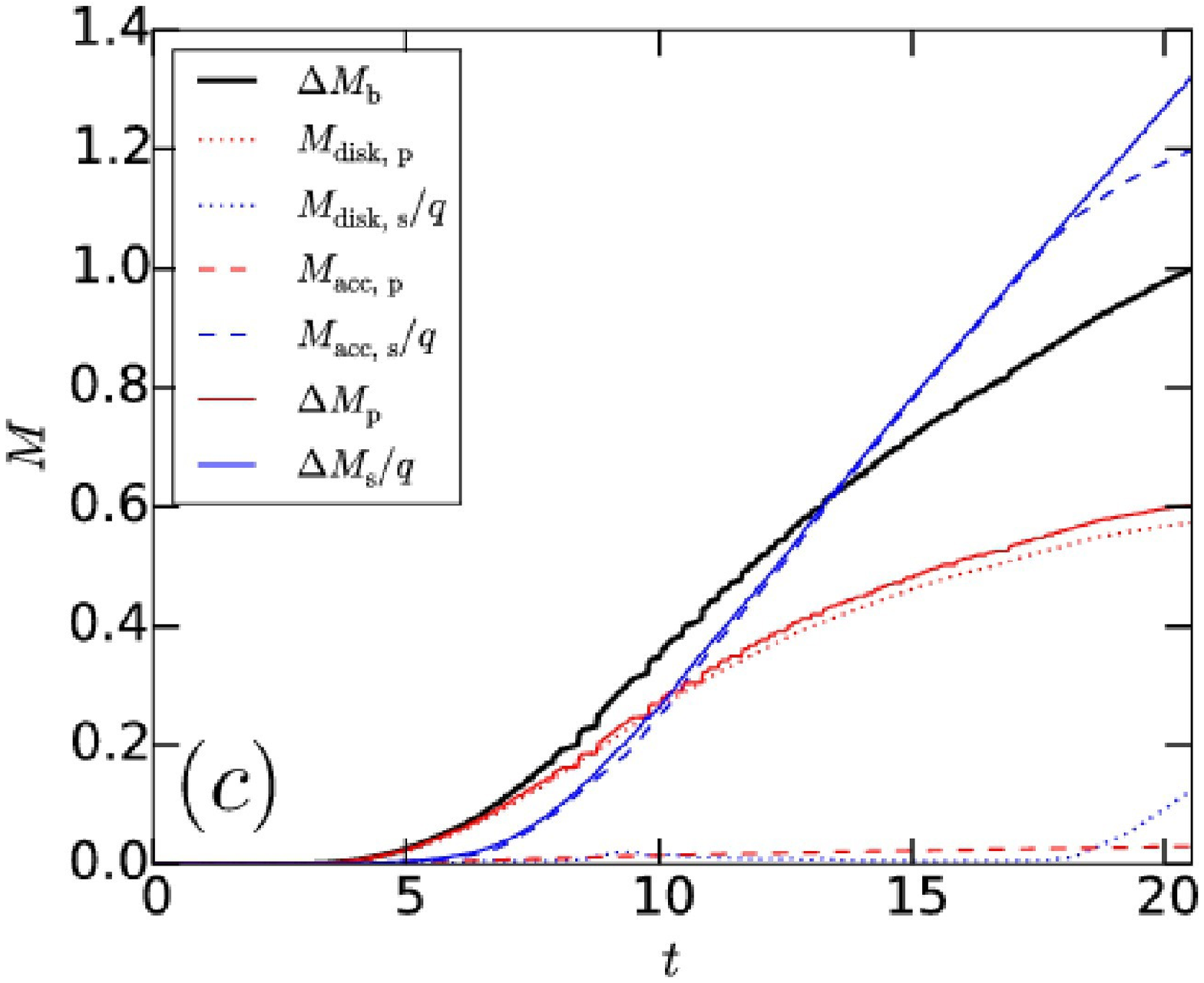}
    \end{center}
  \end{minipage}
  \begin{minipage}{\columnwidth}
    \vspace{-3mm}
    \begin{center}
      \includegraphics[width=\columnwidth]{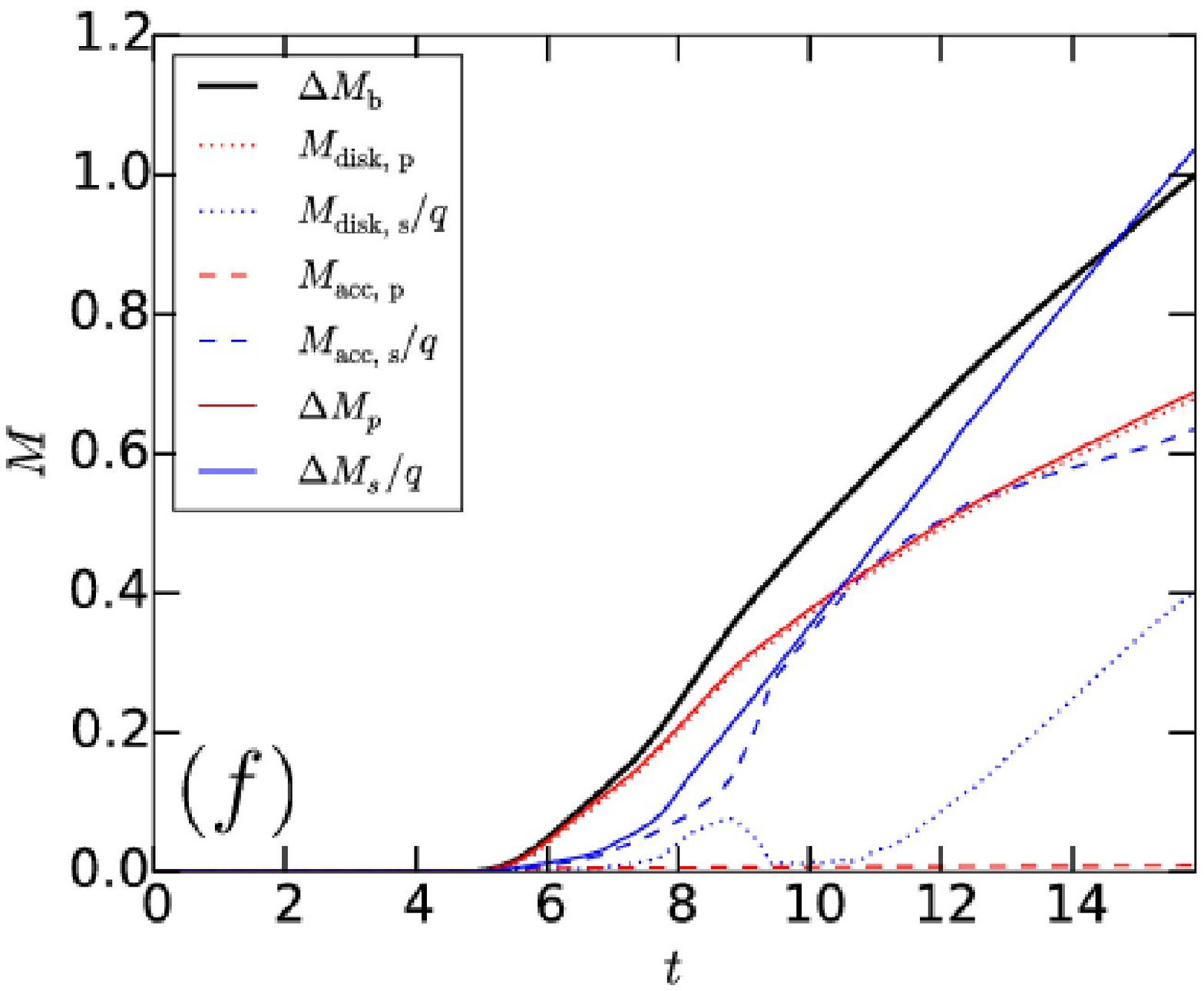}
    \end{center}
  \end{minipage}
  \caption{Same as Fig.~\ref{fig:q01}, but with $q_0=0.3$. Panels ($a$) and ($d$) are at $t=2\pi$, panel ($b$) is at $t=6\pi$, and panel ($e$) is at $t=4\pi$.}
  \label{fig:q03}
\end{figure*}

\begin{figure*}
  \begin{minipage}{\columnwidth}
    \begin{center}
      \includegraphics[width=\columnwidth]{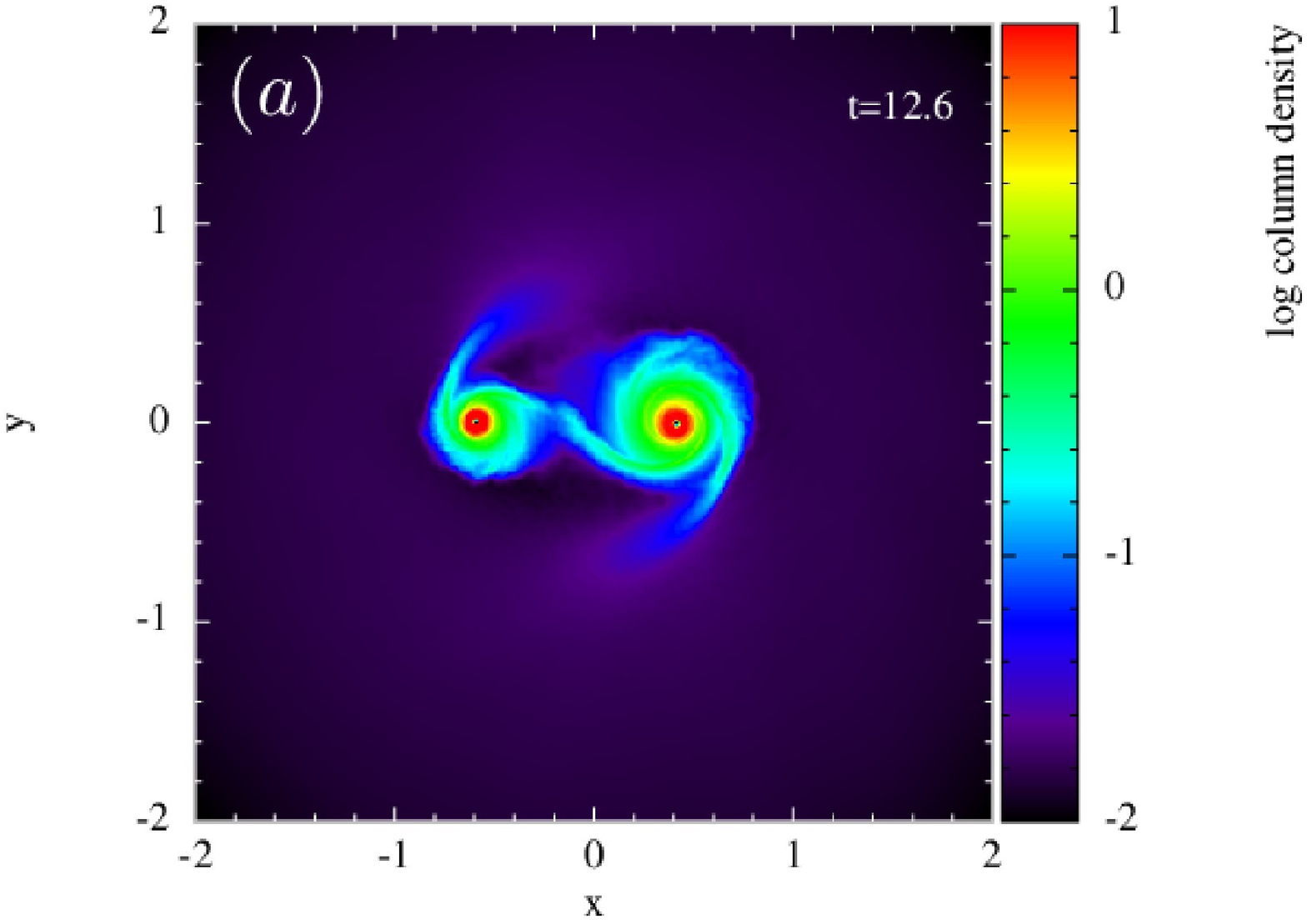}
    \end{center}
  \end{minipage}
  \begin{minipage}{\columnwidth}
    \begin{center}
      \includegraphics[width=\columnwidth]{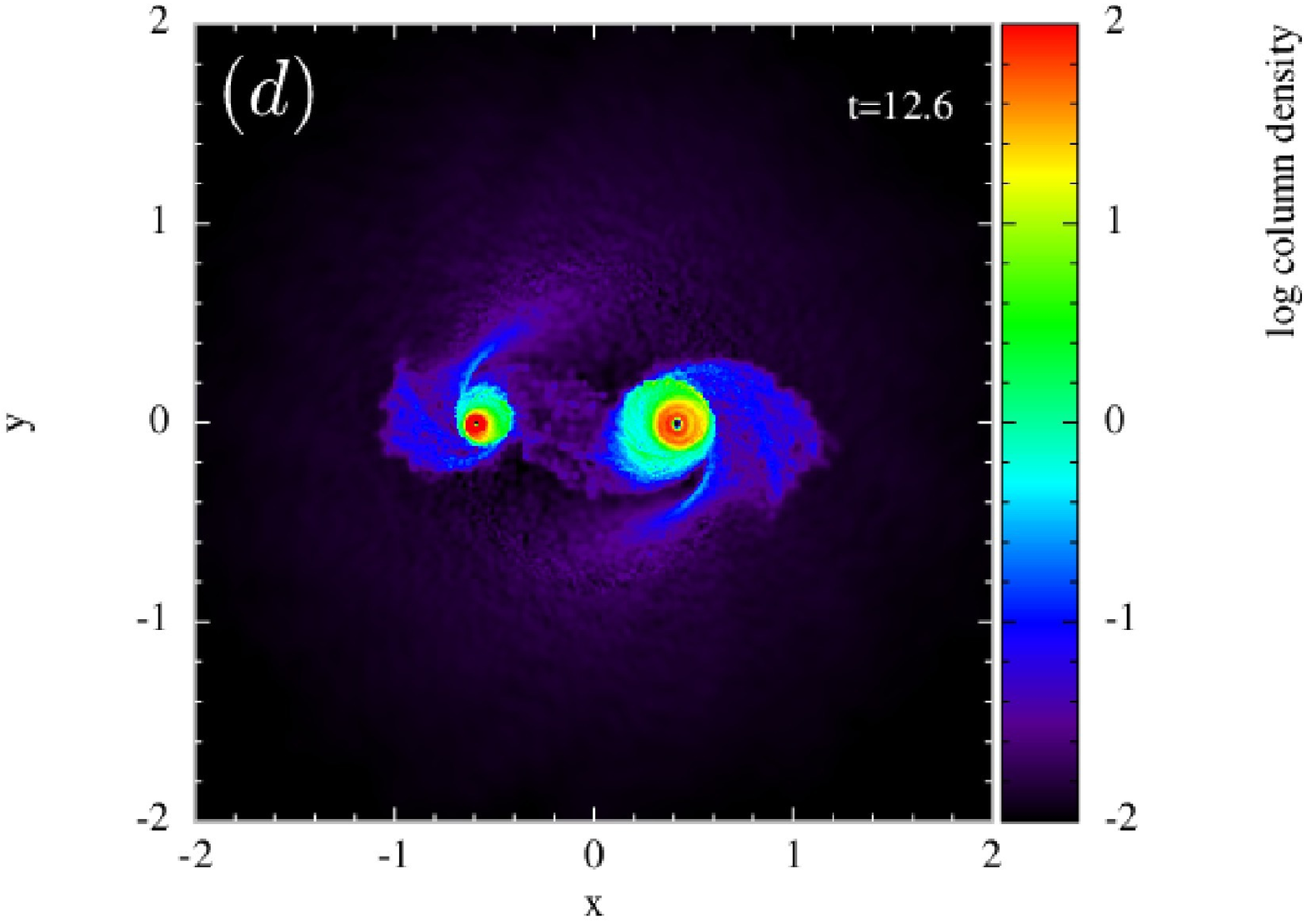}
    \end{center}
  \end{minipage}
  \begin{minipage}{\columnwidth}
    \begin{center}
      \includegraphics[width=\columnwidth]{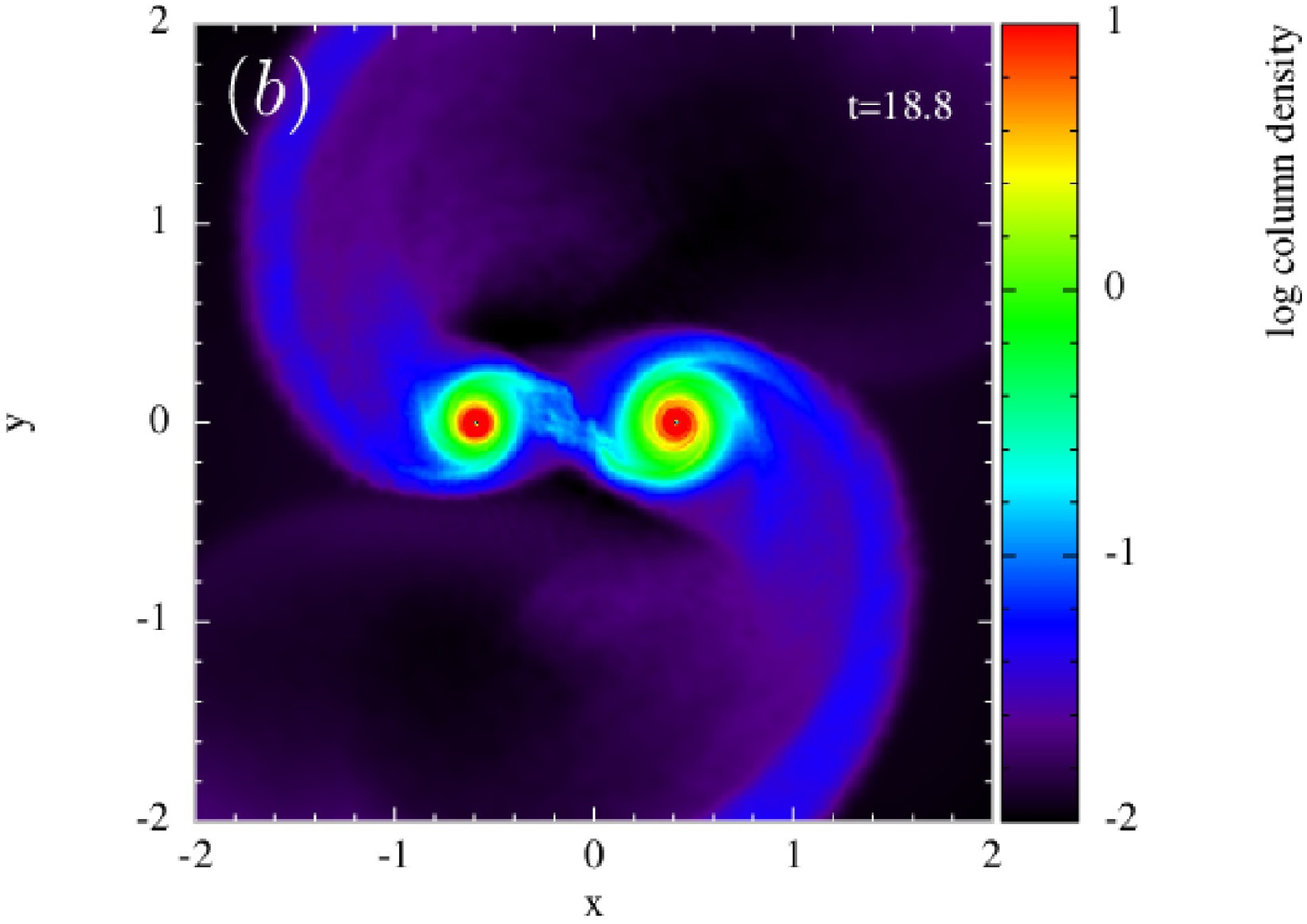}
    \end{center}
  \end{minipage}
  \begin{minipage}{\columnwidth}
    \begin{center}
      \includegraphics[width=\columnwidth]{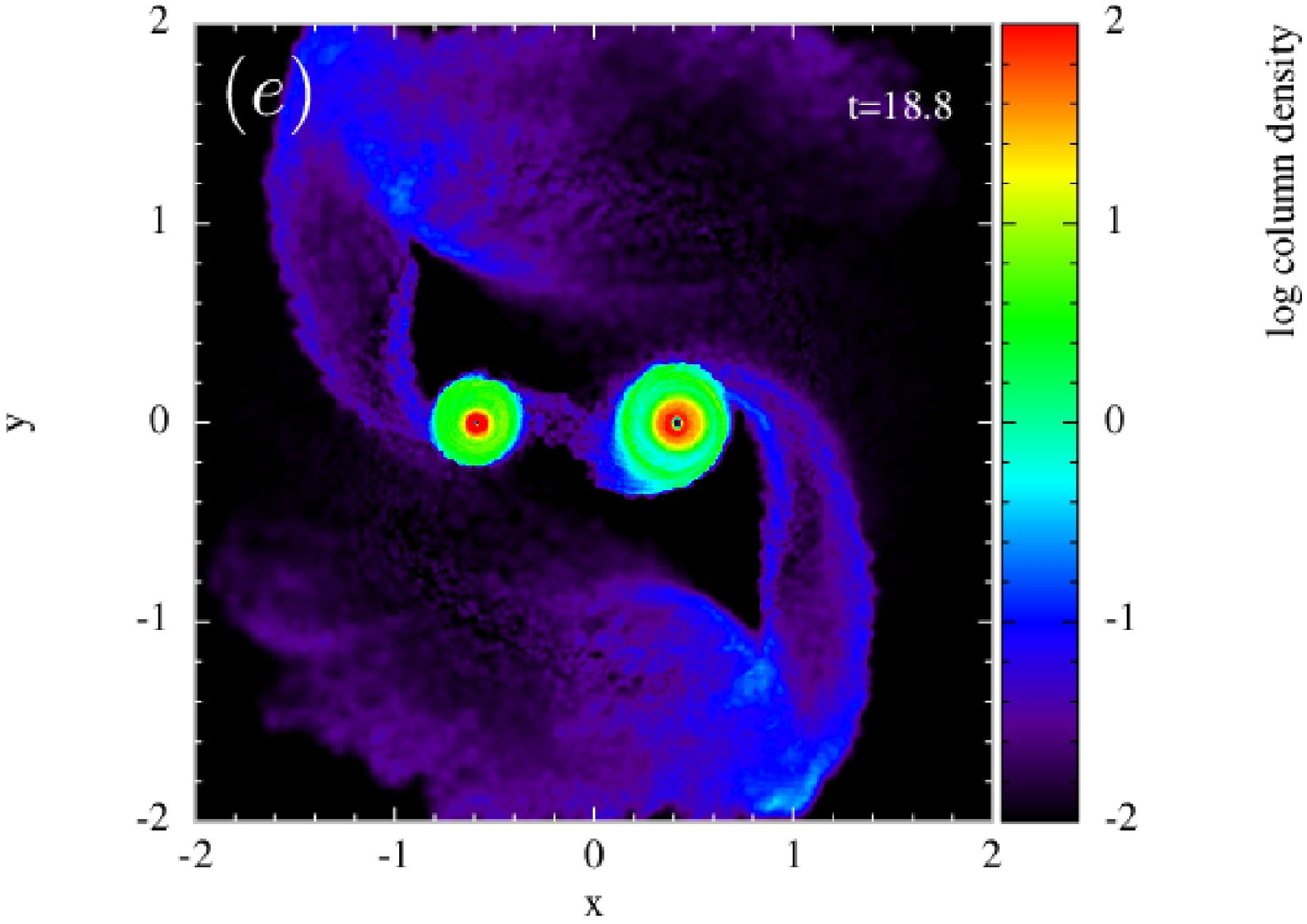}
    \end{center}
  \end{minipage}
  \begin{minipage}{\columnwidth}
    \vspace{-3mm}
    \begin{center}
      \includegraphics[width=\columnwidth]{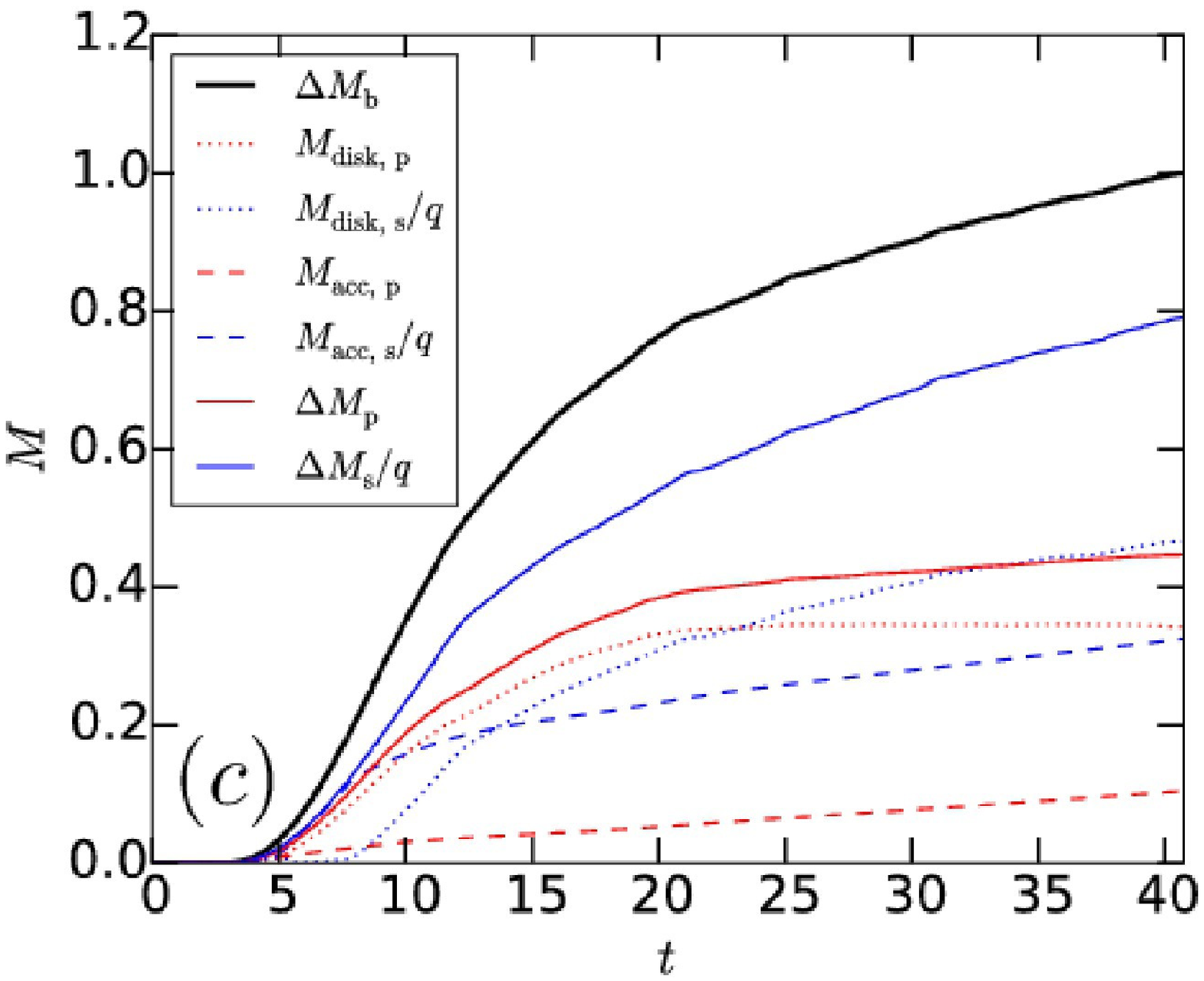}
    \end{center}
  \end{minipage}
  \begin{minipage}{\columnwidth}
    \vspace{-3mm}
    \begin{center}
      \includegraphics[width=\columnwidth]{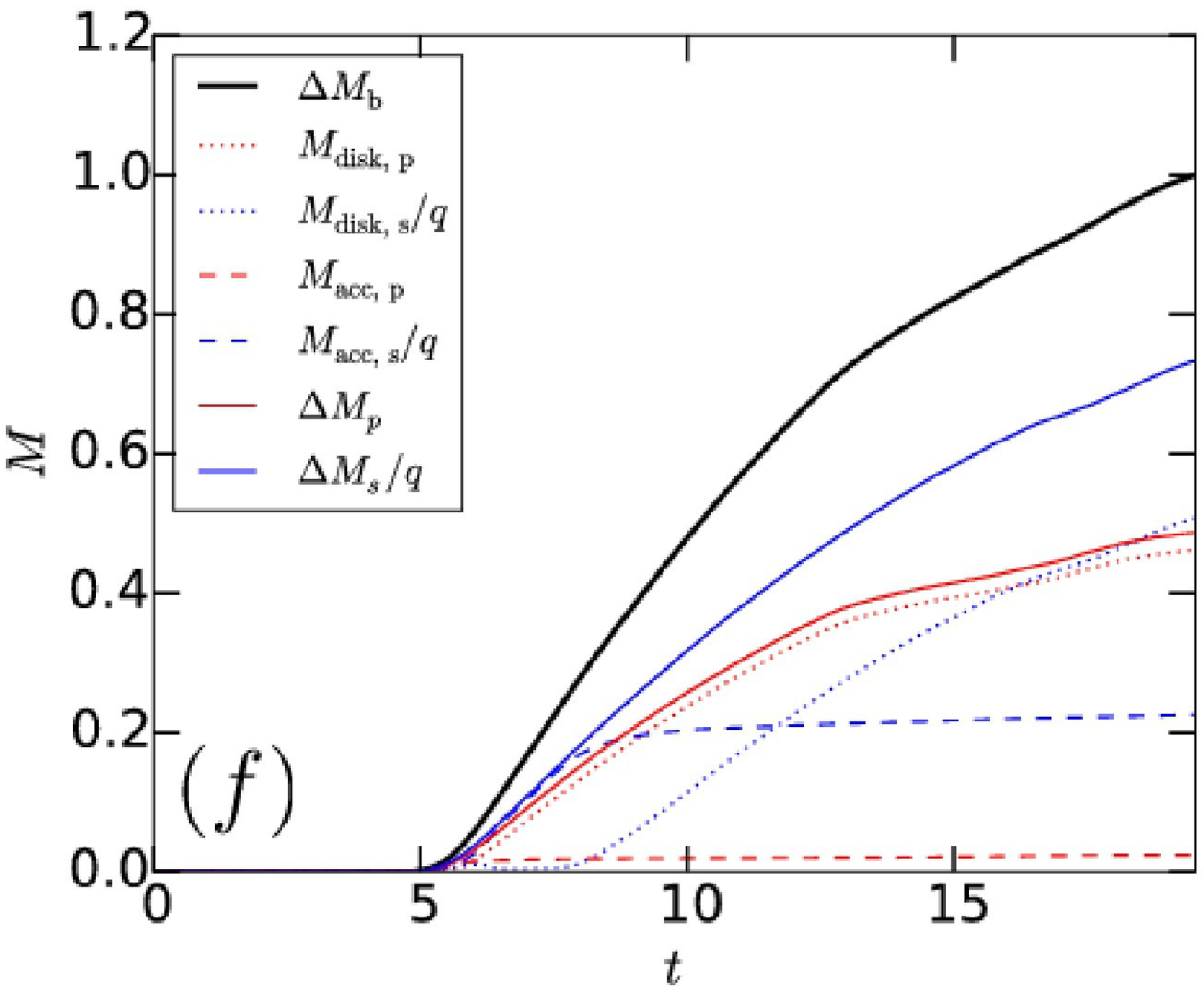}
    \end{center}
  \end{minipage}
  \caption{Same as Fig.~\ref{fig:q01}, but with $q_0=0.7$. Top panels are at $t=4\pi$, and middle panels at $t=6\pi$.}
  \label{fig:q07}
\end{figure*}

In Figure~\ref{fig:q01}, we show the time evolution of gas surface density (top and middle panels),
 and the time evolution of the circum-stellar discs and accreted mass onto each seed
 (bottom panels) for $q_0=0.1$ case. 
For both hot (left column) and cold (right column) cases, 
the circum-primary disc appears at $t \sim 2\pi$ (Fig. \ref{fig:q01}a,d).
At this time, a clear density enhancement with a bridge-like feature exists between the secondary and the primary.
Inside this bridge, we find that the Jacobi constant of gas is dissipated and becomes $J<U_{\rm L1}$, therefore we regard this region as a shock.
Neither the circum-secondary disc nor the circum-binary disc appears until the final state (Fig.~\ref{fig:q01}b,e). 
In Fig.~\ref{fig:q01}c,f, 
it is seen that the red solid line ($\Delta \Mp$) is greater than the blue solid line ($\Delta \Ms/q$). 
This means that the mass ratio decreases, as $\Delta \Mp$ and $\Delta \Ms/q$ are compared here. 
In this case of $q_0=0.1$, we see that the growth of $\Delta \Mb(t)$ is always dominated by that of the circum-primary disc $M_{\rm disc,p}(t)$.
Therefore, the mass ratio monotonically decreases for both hot and cold cases. 

Figure~\ref{fig:q03} is the same as Fig. \ref{fig:q01}, except that it is for $q_0=0.3$.
The circum-primary disc and the shock between the primary and the secondary appear at $t\sim 2\pi$ (Fig.~\ref{fig:q03}a,d), similarly to $q_0=0.1$ case. 
In Fig.~\ref{fig:q03}b,e, the circum-secondary disc appears.
From Fig.~\ref{fig:q03}c,f, it is seen that $\Delta \Mp  < \Delta \Ms/q$ (i.e., mass ratio increases) after $t\sim 3\pi$ for both hot and cold cases. 

Fig.~\ref{fig:q07} is the same as Fig.~\ref{fig:q01} and \ref{fig:q03}, but with $q_0=0.7$. 
In this case, both of the circum-primary disc and the circum-secondary disc are simultaneously formed after $t\sim 2\pi$.
At $t\sim 4\pi$ (Fig.~\ref{fig:q07}a,d), circum-primary disc and circum-secondary disc are clearly seen. 
At $t\sim 6\pi$ (Fig.~\ref{fig:q07}b,e), the circum-binary disc is seen.
After that, gas is accreted onto the circum-binary disc first, before being accreted onto the seeds.  Then the gas arrives at circum-stellar discs through L1 or L2 point.

\begin{figure*}
  \begin{minipage}{\columnwidth}
    \begin{center}
      \includegraphics[width=\columnwidth]{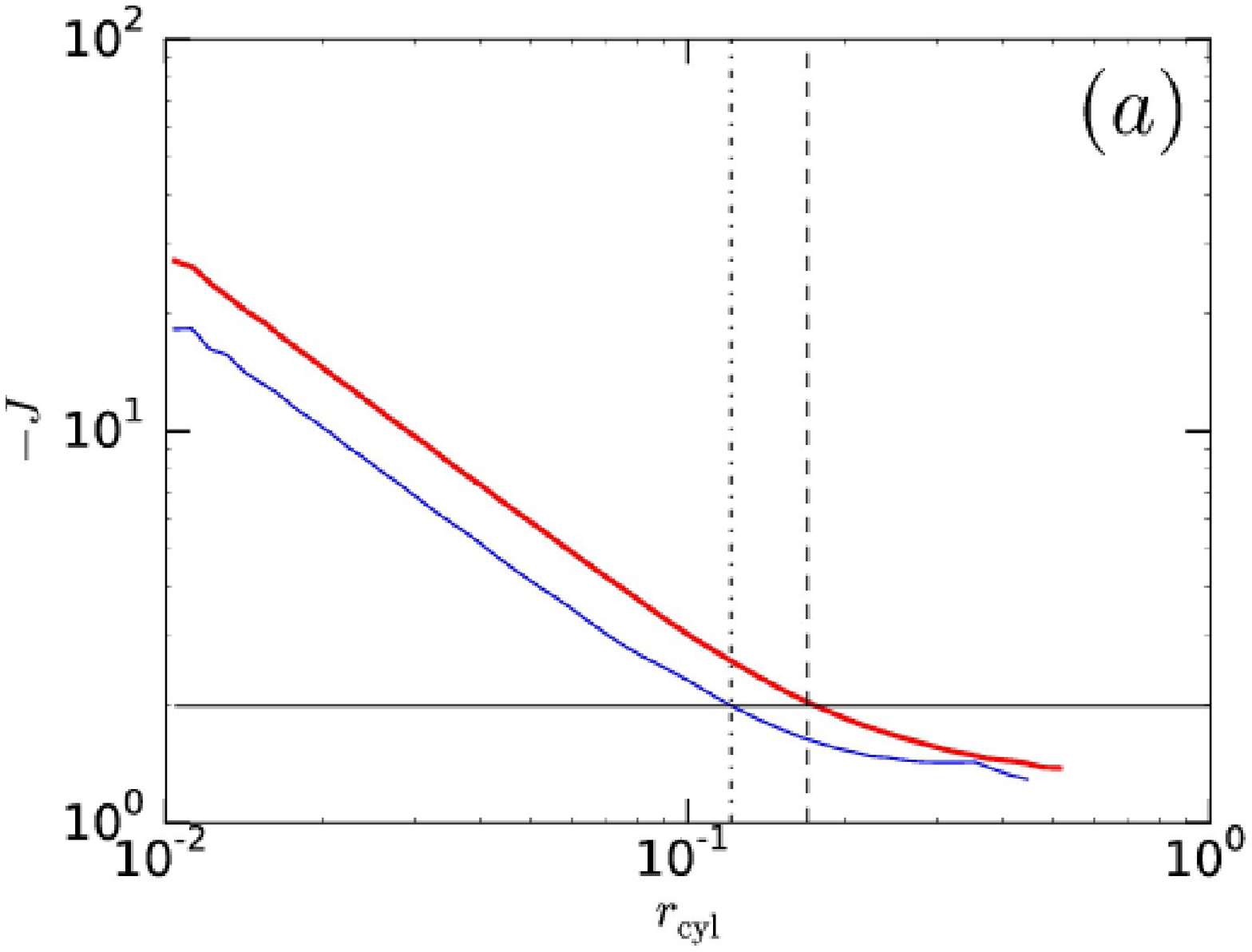}
    \end{center}
  \end{minipage}
  \begin{minipage}{\columnwidth}
    \begin{center}
      \includegraphics[width=\columnwidth]{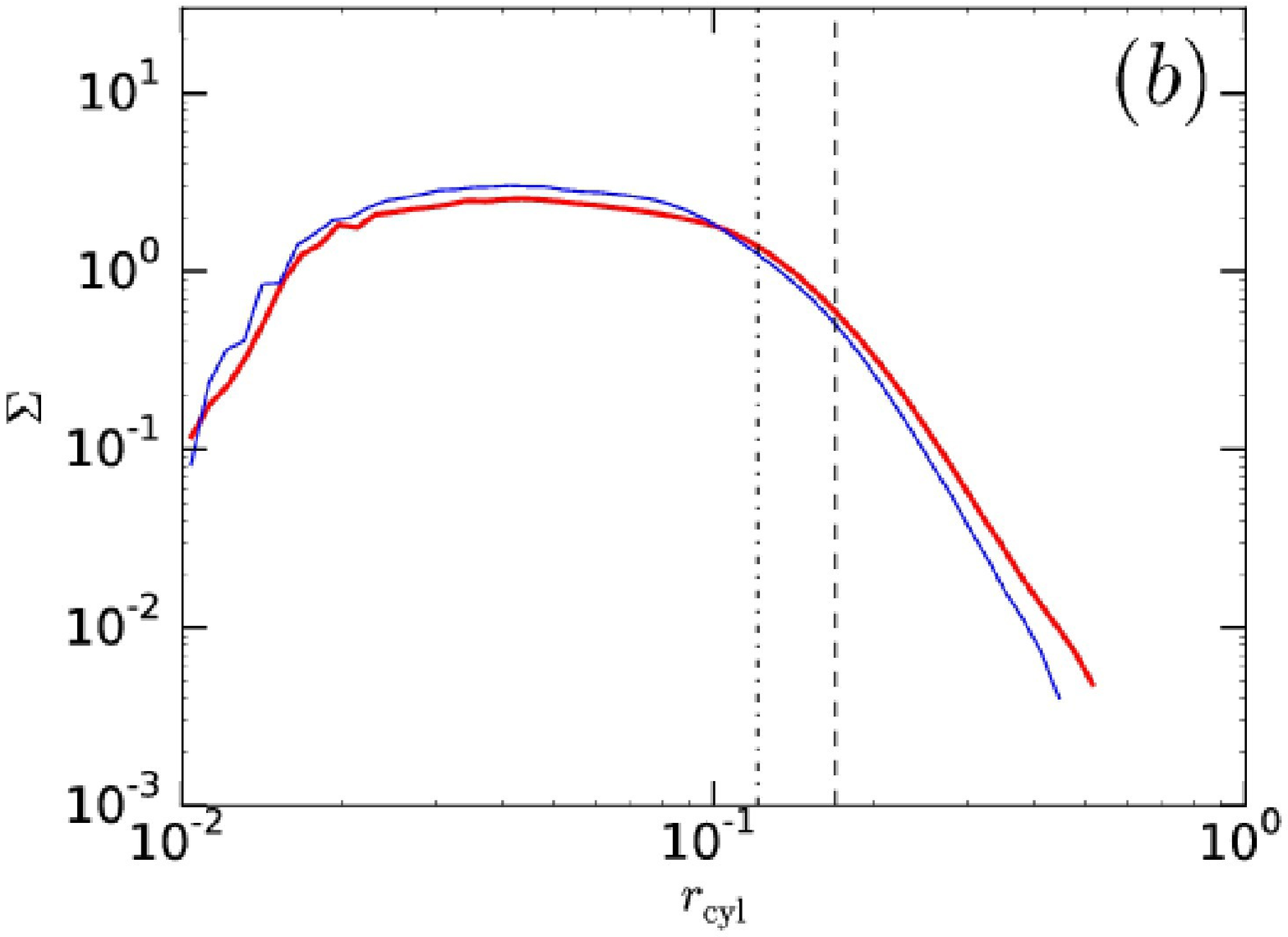}
    \end{center}
  \end{minipage}
  \begin{minipage}{\columnwidth}
    \begin{center}
      \includegraphics[width=\columnwidth]{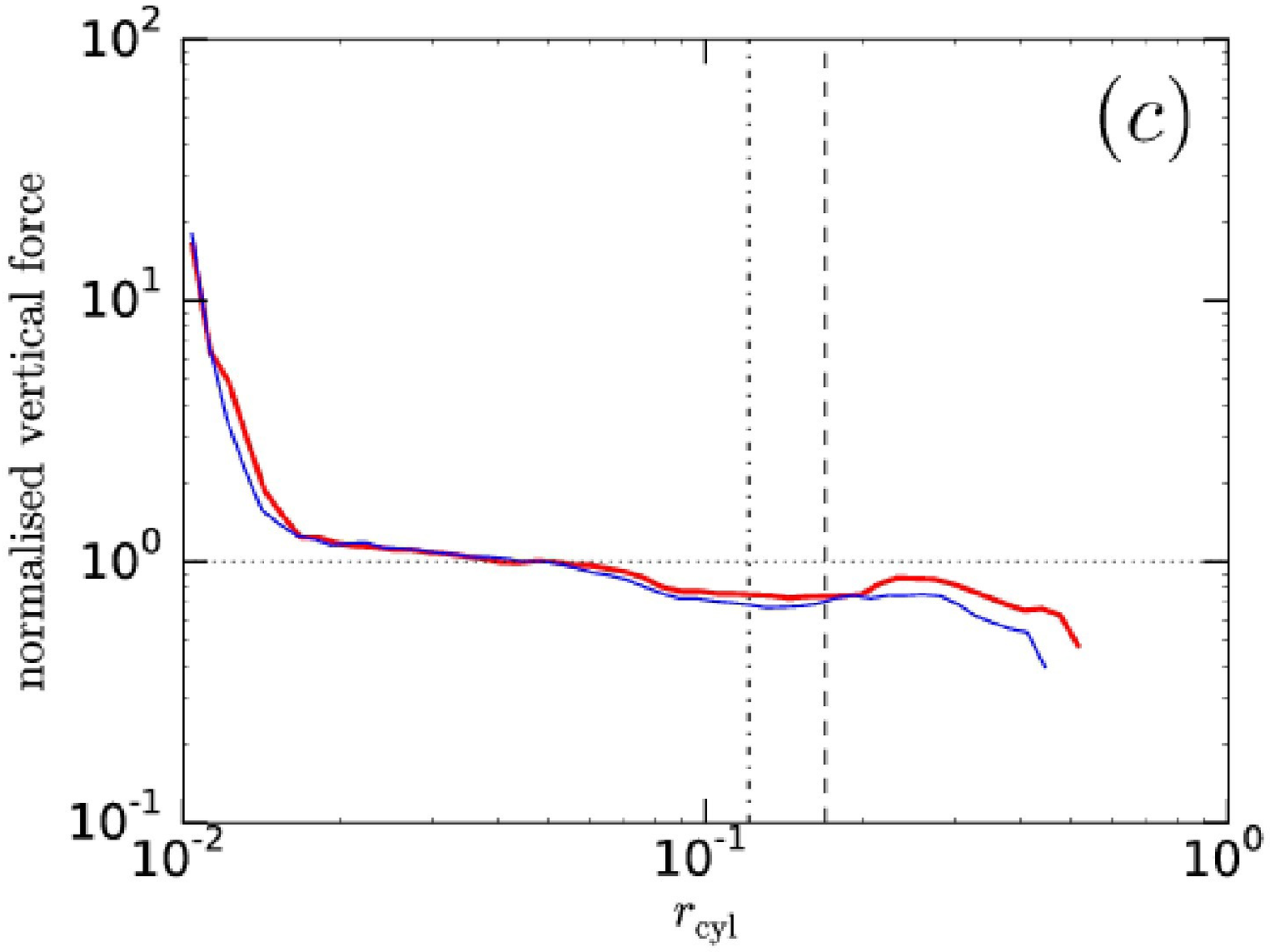}
    \end{center}
  \end{minipage}
  \begin{minipage}{\columnwidth}
    \begin{center}
      \includegraphics[width=\columnwidth]{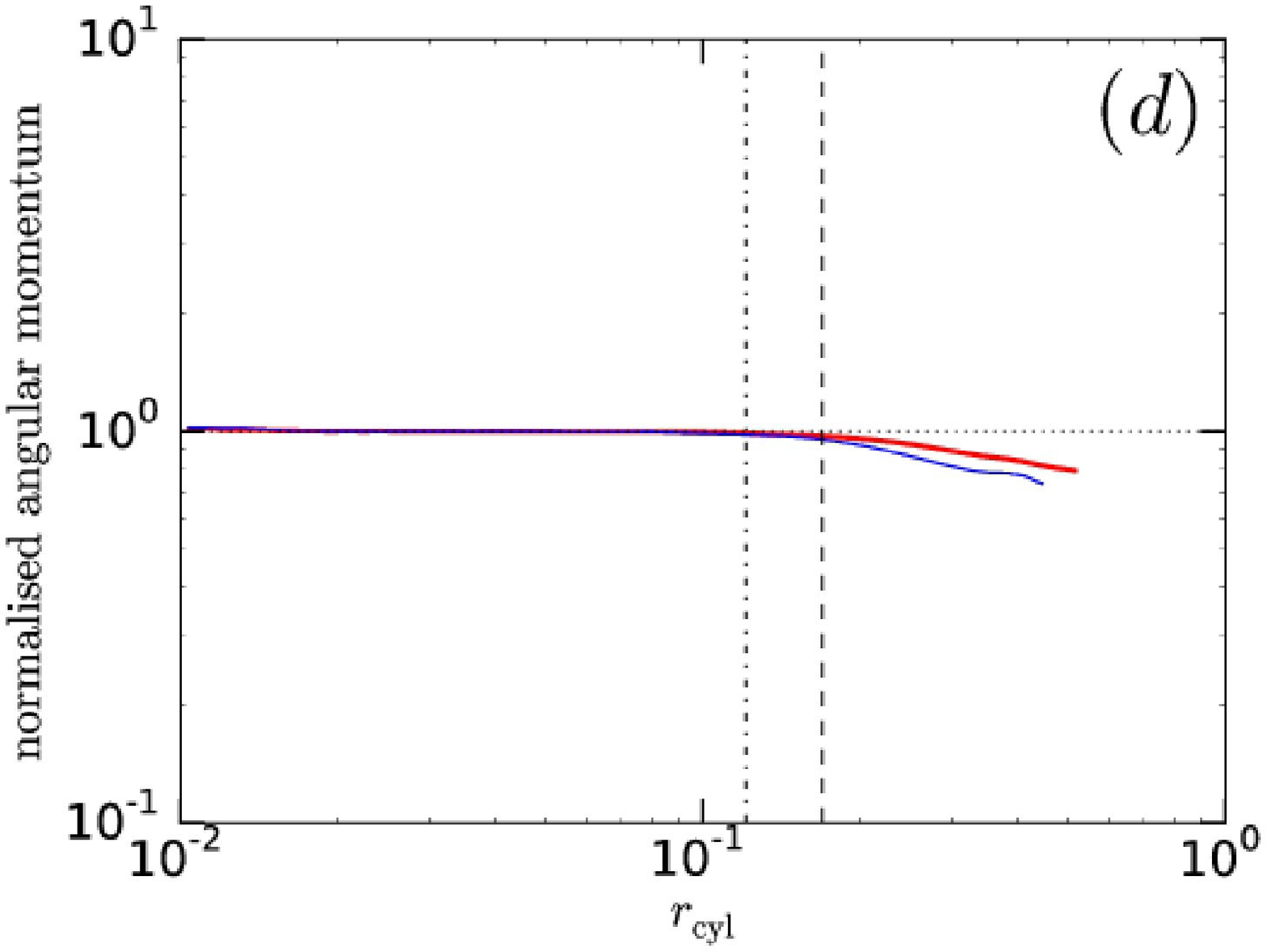}
    \end{center}
  \end{minipage}
  \caption{Properties of circum-stellar discs at the final state of $q_0=0.7$ and $c_s=0.25$ (hot) case. 
In each panel, thick red and thin blue solid lines are for the circum-primary and circum-secondary disc, respectively. 
The vertical black dashed and dot-dashed line denote the edge of circum-primary and circum-secondary disc, respectively. 
The abscissa is the cylindrical radius from each seed.
{\it Panel (a):}  Jacobi constant, with the horizontal black solid line showing $-U_{\rm L1}$. 
{\it Panel (b):}  Gas surface density. 
{\it Panel (c):} Magnitude of vertical gravitational force divided by the magnitude of vertical gas pressure gradient force.  The black horizontal dotted line represents the equilibrium of vertical forces. 
{\it Panel (d):} Specific angular momentum around each seed, normalised by $r_{\rm cyl} v_{\rm K}(1-c_s^2/v_{\rm K}^2)$.}
  \label{fig:DiskValues_hot_q07}
\end{figure*}

Fig.~\ref{fig:DiskValues_hot_q07} shows radial profiles of Jacobi constant (Fig.~\ref{fig:DiskValues_hot_q07}a), gas surface density (Fig.~\ref{fig:DiskValues_hot_q07}b), normalised gravitational force in vertical direction (Fig.~\ref{fig:DiskValues_hot_q07}c), and normalised specific angular momentum (Fig.~\ref{fig:DiskValues_hot_q07}d) of circum-stellar discs at the final state in the hot case with $q_0=0.7$.
In this figure, each distribution is averaged in each cylindrical shell.
In Fig.~\ref{fig:DiskValues_hot_q07}a, the outer edge of the circum-primary disc $R_{\rm disc,p}$ (dashed line) and the outer edge of the circum-secondary disc $R_{\rm disc,s}$ (point-dashed line) are identified by $J=U_{\rm L1}$.
In Fig.~\ref{fig:DiskValues_hot_q07}b, $R_{\rm disc,p}$ (or $R_{\rm disc,s}$) is indeed consistent with the location where the surface density of the circum-primary disc (or the circum-secondary disc) start to decline in the outer region. 
Thus, the circum-stellar discs indeed consists of gas with $J<U_{\rm L1}$.
Another decline in the surface density appears at $r_{\rm cyl} \sim 0.02$, which is caused by the sink particle approximation.
In Fig.~\ref{fig:DiskValues_hot_q07}c, it is seen that the vertical pressure gradient force roughly balances with gravitational force at the region of $0.02<r_{\rm cyl}<R_{\rm disc,p}$ or $R_{\rm disc,s}$, indicating an equilibrium in the vertical direction.
In Fig.~\ref{fig:DiskValues_hot_q07}d, it is seen that angular momentum of gas is approximately Keplerian inside $R_{\rm disc,p}$ or $R_{\rm disc,s}$. 
In summary, the circum-stellar discs consist of gas with $J<U_{\rm L1}$ and are vertically supported by pressure gradient, and rotation velocity of disc is almost Keplerian.
This structure is often seen in the standard disc \citep{Shakura_Sunyaev_73}.


\subsection{Short-term evolution of mass ratio}

\begin{figure}
  \begin{center}
    \includegraphics[scale=0.46]{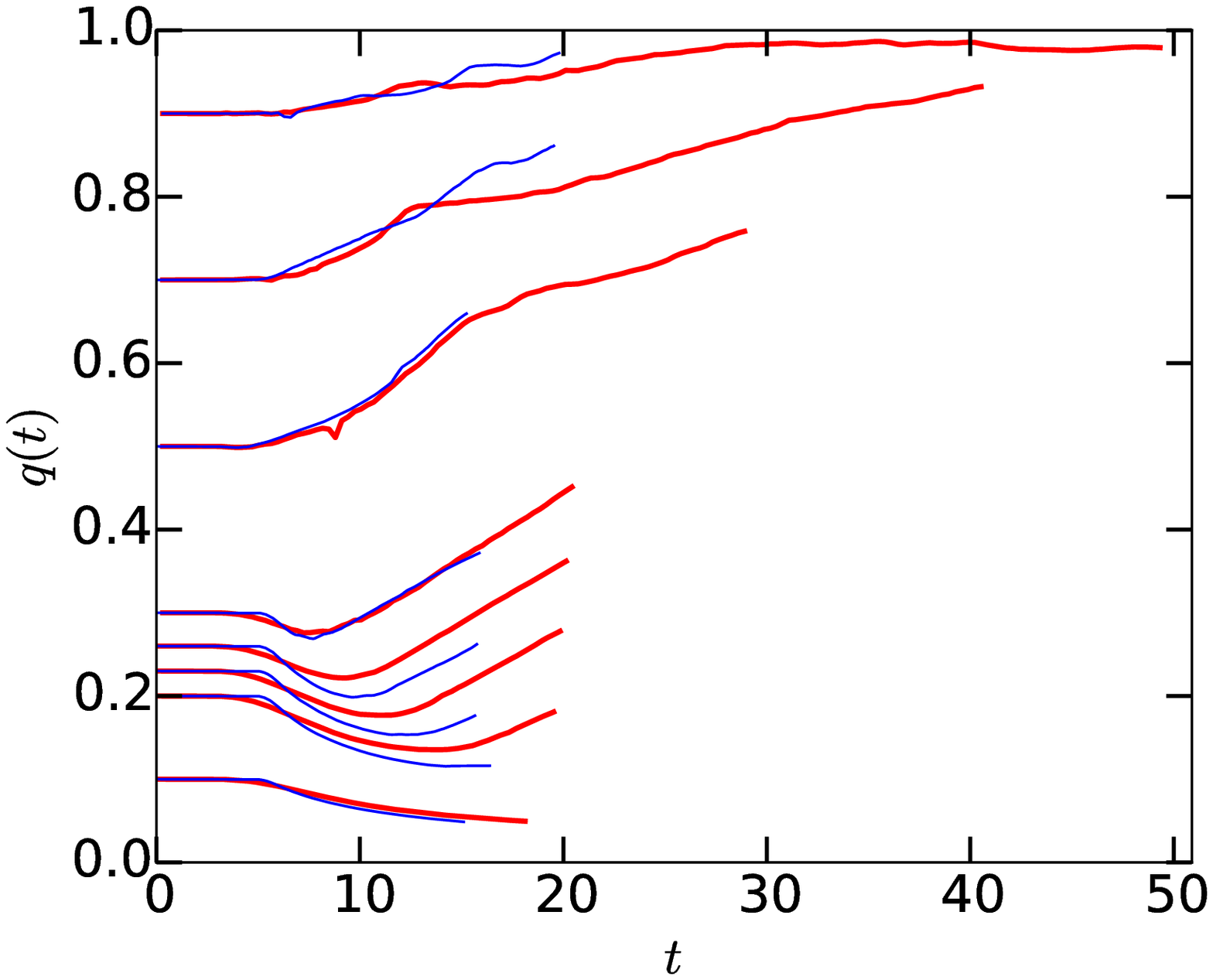}
  \end{center}
  \caption{Time evolution of the mass ratio $q(t)$ in hot (thick red lines) and cold (thin blue lines) cases.
The simulations are terminated when the accreted mass reaches $\Delta \Mb(t) = \Mb$.
}
  \label{fig:Evo}
\end{figure}

Figure~\ref{fig:Evo} shows the time evolution of mass ratio $q(t)$. 
In the cases with $q_0=0.1$, $q(t)$ decreases monotonically.
With $q_0=0.3$, $q(t)$ decreases until $t\sim 3\pi$, 
but it starts to increase after $t\sim 3\pi$, and eventually exceeds $q_0$.
In the cases with $q_0>0.5$, $q(t)$ increases monotonically. 
Taken together, our results show that the time evolution of the mass ratio is qualitatively determined by $q_0$.
If $q_0<0.23$ (in hot case) or $q_0<0.26$ (in cold case), the mass ratio at the final state is smaller than $q_0$. 
If $q_0>0.23$ (in hot case) or $q_0>0.26$ (in cold case), on the other hand, the mass ratio at the final state is larger than $q_0$.
Dependence on temperature clearly appears at $t>4\pi$ in the cases with $q_0>0.7$.
In these cases, the circum-stellar discs settle in a steady state.
\cite{Young_etal_15} found that, in steady circum-stellar discs, hot gas easily crosses the L1 point from the secondary's Roche lobe to the primary's Roche lobe compared to the cold gas, and in such a case the growth of $q(t)$ is suppressed.
In our simulations, the similar tendency as described above is seen in the cases with $q_0>0.7$.

\cite{Bate_00} investigated the evolution of seed binaries with various distribution of angular momentum and density of gas, using the protobinary evolution (PBE) code which employs the steady state solutions of \cite{Bate_Bonnell_97}. In order to examine the results of PBE code, \cite{Bate_00} also performed three dimensional SPH simulations with $N_{\rm SPH}=1\times 10^5$ including angular momentum and density distribution, although limited to one particular case of $q_0=0.6$. 
Although the distributions of $\rho \propto r^{-1}$ and $j\propto r^2$ adopted in \cite{Bate_00} are different from our equations~(\ref{eq:rho_des}) and (\ref{eq:j_des}), our model is similar to one of the models in \cite{Bate_00} because both distributions have the same relation (equation~\ref{eq:rel_j_Mqum}). Thus we can compare our results and that of \cite{Bate_00} except for the time-scale. Focusing on the short-term evolution until $\Delta M_{\rm b} < M_{\rm b}$, our results mentioned in this subsection is consistent with the results of the SPH simulation and the PBE calculations in \cite{Bate_00}. Therefore, our numerical results from SPH simulations confirm the semi-analytical results from PBE calculations in \cite{Bate_00}.

\section{DISCUSSION}\label{sec:discussion}
\subsection{Categorising the Accreting gas}\label{subsec:categorization}
\begin{figure*}
  \begin{center}
       \includegraphics[scale=0.6]{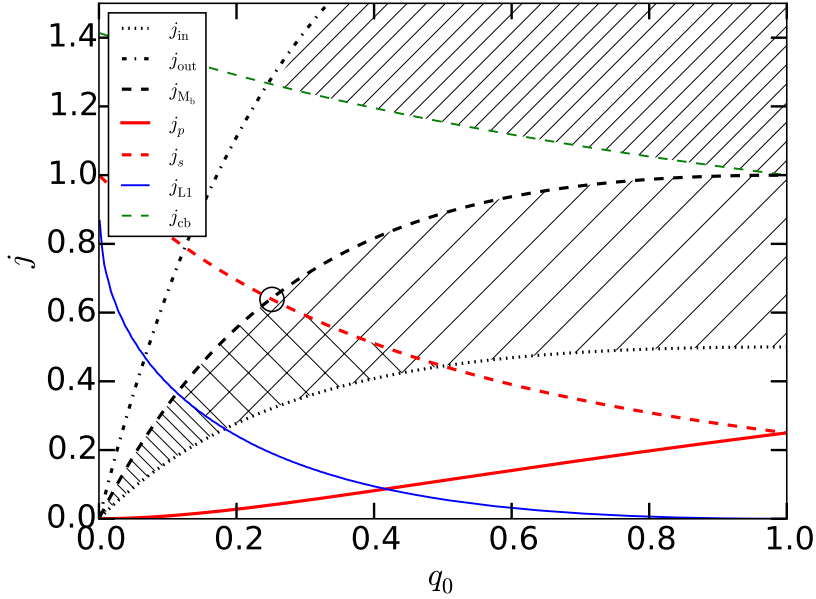}
  \end{center}
  \caption{Relation between the initial mass ratio $q_0$ and the specific angular momentum $j$ of the envelope. Each line shows the specific angular momentum of secondary seed (thick red dashed), primary seed (thick red solid), L1 point (thin blue solid), circum-binary disc (thin green dashed), initial gas specific angular momentum at $R_{\rm in}$ (thick black dotted), at $R_{\rm out}$ (thick-black dot-dashed), and at $r_{\rm Mb}$ (thick black dashed). 
  The black open circle at the intersection of $j_{\rm p}$ and $j_{\rm M_b}$ indicates the critical value $\qc=0.25$. 
 Each shaded region indicates a different mode of gas accretion:  {\it circum-primary disc mode} (backslash), {\it marginal mode} (mesh), {\it circum-stellar discs mode} (single slash), and {\it circum-binary disc mode} (double slash).  
  }
  \label{fig:AngMomCriterion}
\end{figure*}

Focusing on the short-term evolution while $\Delta \Mb(t) \lid \Mb$, 
the accreting gas onto the seed binary can be categorised into four different modes as we describe below.
To characterise the properties of accreting gas in each mode, it is useful to plot the relation 
between initial specific angular momentum of gas and $q_0$. 
In Fig.~\ref{fig:AngMomCriterion}, $j_{\rm in}$ (thick black dotted line) and $j_{\rm out}$ (thick black dot-dashed line)
denotes the initial gas specific angular momentum at $R_{\rm in}$ and $R_{\rm out}$, respectively, and $j_{\rm M_b}$ (thick black dashed line) denotes the initial gas specific angular momentum at $r_{\rm M_b}$,  inside which the gas mass is equal to $\Mb$. 
The specific angular momentum of the secondary and the primary are defined as $j_s$ (thick red dashed line) and $j_p$ (thick red solid line).
The specific angular momentum of L1 point is defined as $j_{\rm L1}$ (blue solid line).
The specific angular momentum of the circum-binary disc $j_{\rm cb}$ (green dashed line) is defined as
\begin{equation}
j_{\rm cb} = \sqrt{ \frac{2}{1+q_0} }j_{\rm circ},
\label{eq:j_cb}
\end{equation}
such that the centrifugal potential ${j_{\rm cb}}^2/2{r_{\rm cyl}}^2$ equals to the gravitational potential $G \Mb /r_{\rm cyl}$ at $r_{\rm cyl}=a_0/(1+q_0)$ which is the distance of secondary from the mass centre \citep{Ochi_etal_05}.
Since the initial specific angular momentum of gas is determined by equation~(\ref{eq:j_des}), 
gas with $j_{\rm in}$ is expected to fall first onto the seed binary.
At the end of the short-term evolution, gas with $j_{\rm M_b}$ is expected to fall onto the circum-stellar discs if we ignore the complex dynamics until the gas falls.
In all our simulations, at the end of the short-term evolution, 
more than $80\%$ of gas in the circum-stellar discs comes from the gas whose initial angular momentum is $j_{\rm in}<j<j_{\rm M_b}$.
Thus $j_{\rm M_b}$ adequately represents the specific angular momentum of accreted gas at the end of the short-term evolution.
Here we define the specific angular momentum of accreted gas as $j_{\rm acc}$, which is in the region filled by backslash, mesh, and single slash in Fig.~\ref{fig:AngMomCriterion}.

The first mode of accreting gas is the ``{\it circum-primary disc mode}" (the region filled by backslash in Fig.~\ref{fig:AngMomCriterion}).
We can see this mode when 
\begin{equation}
j_{\rm acc}<j_{\rm L1}.
\label{eq:type1}
\end{equation}
For example, when $q_0=0.1$, all accreted gas satisfies equation~(\ref{eq:type1}), indicating that the gas easily enters inside L1 point where the Jacobi constant of the gas is dissipated by the shock as discussed in Subsection \ref{subsec:results_disc}.
Since L1 point and mass centre of the seed binary are in Roche lobe of the primary, the gas forms a circum-primary disc (Fig.~\ref{fig:q01}b,e), the primary seed grows, and the mass ratio decreases monotonically (Fig.~\ref{fig:Evo}).

The second mode of accreting gas is the ``{\it marginal mode}" (the region filled by mesh in Fig.~\ref{fig:AngMomCriterion}).
This mode is seen when 
\begin{equation}
j_{\rm L1} < j_{\rm acc} < j_{\rm s}.
\label{eq:type2}
\end{equation}
When $q_0=0.2$, for example, all accreted gas satisfies equation~(\ref{eq:type2}). 
In this case, most of gas is trapped by the primary 
similarly to the {\it circum-primary disc mode}.
In the end, the mass ratio decreases.
However, the gas that satisfies equation~(\ref{eq:type2}) enters inside secondary's Roche lobe more easily than in the {\it circum-primary disc mode}. 
Inside secondary's Roche lobe, the Jacobi constant of gas is dissipated by the shock.
As a result,  $M_{\rm acc,s}$ becomes non-negligible in the end.

The third mode of accreting gas is ``{\it circum-stellar discs mode}" (the region filled by single slash in Fig.~\ref{fig:AngMomCriterion}).
We can see this mode when 
\begin{equation}
j_{\rm s} < j_{\rm acc} < j_{\rm cb}.
\label{eq:type3}
\end{equation}
In this case, a circum-secondary disc is formed.
Once the circum-secondary disc is formed, $\Delta M_{\rm s}/q_0$ dominates, 
and the mass ratio increases monotonically.
We can see this mode when $q_0=0.7$, for example (see Fig.~\ref{fig:q07}).

The fourth mode is the ``{\it circum-binary disc mode}" (the region filled by double slash in Fig.~\ref{fig:AngMomCriterion}).
We can see this mode when the specific angular momentum of gas is larger than $j_{\rm cb}$ (equation~\ref{eq:j_cb}),
\begin{equation}
j > j_{\rm cb}. 
\end{equation}
In this mode, the majority of gas cannot enter inside each Roche lobe because of the centrifugal barrier, and gas settles down to the circum-binary disc first. 
Then, the gas enters inside each Roche lobe through L2 or L3 point, and falls onto the circum-stellar discs.
This behaviour is seen at $t>6\pi$ for $q_0=0.7$ (see Fig.~\ref{fig:q07}b,e).
Since $j_{\rm M_b}$ is lower than $j_{\rm cb}$ for any $q_0$ in our simulations (Fig.~\ref{fig:AngMomCriterion}), the gas in {\it circum-binary disc mode} is not accreted by the end of short-term evolution.
Therefore, the {\it circum-binary disc mode} is irrelevant for the $q$-evolution in the short term.
To investigate $q$-evolution in this mode, we need to simulate the long-term evolution.

In our simulations, the time evolution of mass ratio qualitatively changes at $q_{\rm c,hot}=0.23$ (hot case) or $q_{\rm c,cold}=0.26$ (cold case).
The values of $q_{\rm c,hot}$ and $q_{\rm c,cold}$ roughly correspond to the intersection point of $j_{\rm s}$ and $j_{\rm {M_b}}$ in Fig.~\ref{fig:AngMomCriterion}.
Therefore, we define a critical initial mass ratio $\qc$ at this intersection point, 
and we find $\qc=0.25$ from Fig.~\ref{fig:AngMomCriterion}.
The value of $q_{\rm c,cold}$ is somewhat closer to $\qc$ than $q_{\rm c,hot}$.
This is because the gas flow is closer to a ballistic motion in the cold limit than in the hot case. 
With a finite gas temperature, pressure gradient force pushes out the gas in radial direction.
Therefore, even if $j_{\rm M_b}<j_{\rm L1}$, the rotation radius of gas with $j_{\rm M_b}$ can reach ${j_{\rm L1}}^2/G \Mb$.
Since $j_{\rm M_b}$ is monotonically increasing function of $q_0$, $q_{\rm c,hot}$ is somewhat lower than $q_{\rm c}$. 
The difference between $q_{\rm c,hot}$ and $q_{\rm c,cold}$ is small since this push-out effect is expected to be weak when $c_s/v_{\rm K}<1$.
Here we emphasize that the critical value $\qc=0.25$ was derived only for a particular distribution of angular momentum and density (equation~\ref{eq:rel_j_Mqum}), and that it was evaluated when $\Delta M_{\rm b}(t) = M_{\rm b}$.

In summary,  gas accretion onto the primary dominates in 
the {\it circum-primary disc mode} and the {\it marginal mode}.
While in the {\it circum-stellar discs mode}, a circum-secondary disc is formed and accretion onto the secondary becomes significant enough to increase the mass ratio. 
The gas in {\it circum-binary disc mode} forms a circum-binary disc.

\subsection{Analytic Estimate of Long-term Evolution}\label{subsec:analytic}
In our numerical simulations, we focus on the short-term evolution until $\Delta \Mb(t) = \Mb$ assuming an isolated binary with no self-gravity. 
In this subsection, we discuss the long-term evolution of binary separation analytically including binary growth by accretion.

There are two effects which change the binary separation by accretion.
One is the increase of binary mass. 
When the binary mass becomes larger and if the angular momentum is conserved, 
then the binary separation becomes smaller because of stronger gravitational force. 
The other is the increase of binary angular momentum, which increases the binary separation. 
The evolution of binary separation is determined by the competition between above two effects.
These effects become especially important when $\Delta \Mb(t)>\Mb$. 
First, we formulate the time evolution of binary in our model.
Then, we discuss one possibility in which the long-term evolution can be predicted based on our numerical results of short-term evolution.

As for the binary, we define the time-dependent binary mass $\Mb(t)$, binary separation $a(t)$, mass ratio $q(t)$. 
The reference specific orbital angular momentum can be written as 
\begin{equation}
j_{\rm circ}(t)=\sqrt{G\Mb(t)a(t)} \label{eq:j_circ(t)}.
\end{equation}
Then the time-dependent orbital angular momentum of binary $J_{\rm b}(t)$ is written by
\begin{equation}
J_{\rm b}(t) = \frac{2q(t)}{(1+q(t))^2} \Mb(t) j_{\rm circ}(t).
\label{eq:J_b(t)}
\end{equation}
We introduce following non-dimensional variables:
\begin{eqnarray}
\tilde{M}(t) &=& \frac{M_{\rm b}(t)}{M_{\rm b}} \label{eq:tilde_M},\\
\tilde{J}(t) &=& \frac{J_{\rm b}(t)}{J_{\rm b}} \label{eq:tilde_J},\\
\tilde{a}(t) &=& \frac{a(t)}{a_0} \label{eq:tilde_a},
\end{eqnarray}
and $j_{\rm circ}(t)$ is represented as
\begin{equation}
j_{\rm circ}(t)=\frac{(1+q(t))^2}{q(t)}\frac{q_0}{(1+q_0)^2}\frac{{\tilde J}(t)}{{\tilde M}(t)}j_{\rm circ}.
\end{equation}
Note that we stop our simulations when it becomes $\tilde{M}=2$. 

As for the envelope, in our model (equations~\ref{eq:rho_des} and \ref{eq:j_des}), 
the specific angular momentum of gas $j$ and the gas mass inside the radius $r$, $M_{\rm gas}$, 
has a relationship
\begin{equation}
j\propto M_{\rm gas} \propto r.
\label{eq:rel_j_Mqum}
\end{equation}
From equations~(\ref{eq:tilde_M}) and (\ref{eq:rel_j_Mqum}), 
$j_{\rm in}$ as a function of time is given by
\begin{equation}
j_{\rm in}(t) = \tilde{M}j_{\rm in} \label{eq:j_in(t)}.
\end{equation}
From equations~(\ref{eq:j_0}), ~(\ref{eq:j_circ(t)}) and (\ref{eq:j_in(t)}), we have
\begin{eqnarray}
j_{\rm in} &=& \frac{2q_0}{(1+q_0)^2}j_{\rm circ},\label{eq:j_0_again}\\
j_{\rm in}(t) &=& \frac{2q(t)}{(1+q(t))^2}\frac{{\tilde M}^2(t)}{{\tilde J(t)}}j_{\rm circ}(t).
\label{eq:j_0_t}
\end{eqnarray}
Equations~(\ref{eq:j_0_again}) and (\ref{eq:j_0_t}) represent the specific angular momentum at the inner edge of the envelope. 
The power indices of ${\tilde M}$ and ${\tilde J}$ in equation~(\ref{eq:j_0_t}) reflect the spatial distribution of density and angular momentum in the envelope.
If the relation 
\begin{equation}
\frac{{\tilde M}^2(t)}{{\tilde J}(t)}=1\label{eq:self_similar}
\end{equation}
holds and if $q(t)=q_0$, equations~(\ref{eq:j_0_again}) and (\ref{eq:j_0_t}) are the same in units of $M_{\rm b}(t)=a(t)=1$ and $M_{\rm b}=a_0=1$.
This indicates that the evolution of binary system is self-similar when equation~(\ref{eq:self_similar}) holds and $q(t)=q_0$.
Note that, in equations~(\ref{eq:j_in(t)}) and (\ref{eq:j_0_t}), it is implicitly assumed that all angular momentum and mass of the envelope is converted to the orbital angular momentum and mass of the binary.

After the above preparation, we can now discuss the time evolution of binary separation. 
From equations~(\ref{eq:J_b}) and (\ref{eq:J_b(t)}), we have
\begin{equation}
\tilde{a}(t) = \left(\frac{q(t)}{q_0}\right)^{-2} \left(\frac{1+q(t)}{1+q_0} \right)^{4}\frac{{\tilde J}^2(t)}{{\tilde M}^3(t)}.
\label{eq:a_evo}
\end{equation}
From equation~(\ref{eq:a_evo}), we can see that the separation becomes larger with increasing orbital angular momentum of the binary, and that it becomes smaller with increasing mass.
Moreover, the separation also depends on $q(t)$, and this dependence originates from equation~(\ref{eq:J_b(t)}). 
For given $J_{\rm b}(t)$ and $M_{\rm b}(t)$, one can see from equation~(\ref{eq:J_b(t)}) that $a(t)$ inside $j_{\rm circ}$ depends on $q(t)$. 
If equation~(\ref{eq:self_similar}) and $q(t)=q_0$ hold, binary separation is proportional to accreted mass in our model:
\begin{equation} 
\tilde{a}(t) = {\tilde M}(t).\label{eq:a_self}
\end{equation}
The analytic result of equation~(\ref{eq:a_self}) is consistent with the numerical work by \cite{Bate_00}.
Here, we discuss one possibility in which the long-term evolution can be predicted by reusing the result of the short-term evolution.
From equations~(\ref{eq:j_0_again}) and (\ref{eq:j_0_t}), we see that the difference between $j_{\rm in}/j_{\rm circ}$ and $j_{\rm in}(t)/j_{\rm circ}(t)$ is caused only by the mass ratio, if equation~(\ref{eq:self_similar}) always holds. 
According to our simulations, in the hot case with $q_0=0.5$, $q \approx 0.7$ when $\tilde{M}=2$ from Fig.~\ref{fig:Evo}.
Under the above assumptions, we can reuse the former result to predict that 
the mass ratio would be $q \approx 0.9$ when it reaches $\tilde{M}=3$.  
Repeating this procedure, we can predict the long-term evolution of a seed binary. 

\begin{figure}
  \begin{center}
    \includegraphics[width=\columnwidth]{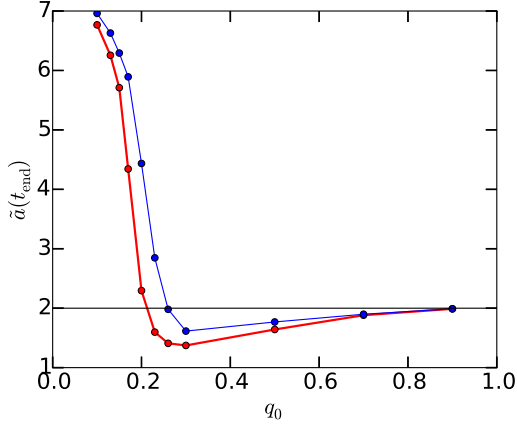}
  \end{center}
  \caption{Binary separation at the end of the short-term evolution $\tilde{a}(t_{\rm end})$ in the cases of hot (thick red line) and cold (thin blue line).  The black horizontal line denotes $\tilde{a}(t_{\rm end}) = 2$.}
  \label{fig:a_evo}
\end{figure}

We saw in Fig.~\ref{fig:Evo} that, in the short-term evolution, $q(t)$ increases monotonically if $q_0>\qc$, and vice versa. 
Based on this result and the argument in the previous paragraph, 
we argue that the long-term evolution of $q(t)$ is qualitatively determined by $q_0$.
Fig.~\ref{fig:a_evo} plots equation~(\ref{eq:a_evo}) at the end of the short-term evolution 
(i.e., binary separation at $\tilde{M} = 2$) using our numerical results of $q(t)$ and equation~(\ref{eq:self_similar}).
Fig.~\ref{fig:a_evo} shows that the separation reaches $\tilde{a}(t_{\rm end})=2$ in the cases with $q_0 \rightarrow 1.0$ and $q_0\simeq q_{\rm c}$, indicating that the time evolution of a binary is self-similar in these cases (equation~\ref{eq:a_self}).
Fig.~\ref{fig:a_evo} also shows that $\tilde{a}(t_{\rm end})>1$ for any $q_0$, which suggests that the binary separation is a monotonically increasing function of time and therefore close binaries are difficult to form.

Here, we note again that these analytic results are based on the assumption that all angular momentum of the envelope is converted to the orbital angular momentum of the binary.
In other words, we are disregarding the division of gas angular momentum into orbital angular momentum of binary and that of circum-stellar discs.
In order to investigate the growth of separation more properly, a direct calculation of the binary orbit is needed.

\section{CONCLUSION AND FUTURE WORK}\label{sec:conclusion}
In the present work, we investigate the short-term evolution of a seed binary using the SPH code {\tt GADGET-3} in three dimensions. 
Our simulation setup includes non-uniform distribution of gas and angular momentum with 
$\rho \propto r^{-2}$ and $j \propto r$, respectively. 
In the initial condition, the seed binary is assumed to have formed around the mass centre of the binary by fragmentation, conserving angular momentum and mass.
The seed binary is isolated, and self-gravity of gas is ignored.
With this setup, we compute the accretion of gaseous envelope onto the seed binary 
until the binary mass growth exceeds its initial mass, surveying the parameter ranges of 
 $0.1<q_0<1.0$ and the sound speeds $c_s/\sqrt{GM_{\rm b}/a_0} = 0.05$ (cold) and $0.25$ (hot). 

As a result, we categorise the gas accretion into four different modes as follows: 
\begin{enumerate}
\item {\it ``Circum-primary disc mode"} is seen when the specific angular momentum of accreting gas is lower than that of L1 point, i.e., $j_{\rm acc} < j_{\rm L1}$.
Most of the gas falls onto the primary and the circum-primary disc, and hence $q(t)$ monotonically decreases. 
This is because the specific angular momentum is small enough, and the gas with $j_{\rm acc} < j_{\rm L1}$ enters the primary's Roche lobe.  
\item {\it ``Marginal mode"} is seen when $j_{\rm L1}<j_{\rm acc}<j_{\rm s}$.
In this case, although most of the gas is trapped by the primary similarly to the {\it "circum-primary disc mode"}, the gas is able to enter the secondary's Roche lobe, and the secondary starts to accrete gas.  As a result $q(t)$ becomes smaller than $q_0$ after the short-term evolution. 
\item {\it ``Circum-stellar discs mode"} is seen when $j_{\rm s} < j_{\rm acc} < j_{\rm cb}$.
If the specific angular momentum of gas exceeds that of the secondary,  gas starts to rotate around the secondary, 
and a circum-secondary disc is also formed.
Once the circum-secondary disc is formed, $q(t)$ monotonically increases. 
\item {\it ``Circum-binary disc mode"} is seen when $j_{\rm cb} < j$.
In this case, gas cannot fall onto the circum-stellar discs directly because of its large angular momentum.
Therefore, the gas falls onto the circum-binary disc first, and then later enter the Roche lobes through L2 or L3 point.
\end{enumerate}

We find that the short-term evolution of $q$-value is qualitatively different according to its initial value $q_0$. 
If $q_0> \qc = 0.25$, the final mass ratio exceeds $q_0$.
This critical value $\qc$ is determined by the condition $j_{\rm s} = j_{\rm M_{\rm b}}$ in Fig.~\ref{fig:AngMomCriterion}. 
The critical value $\qc=0.25$ was derived only for a particular distribution of angular momentum and density (equation~\ref{eq:rel_j_Mqum}), and that it was evaluated when $\Delta M_{\rm b} = M_{\rm b}$.
In {\it circum-primary disc mode} or {\it marginal mode}, 
the dominant accretion onto the primary decreases the $q$-value. 
However, once the circum-secondary disc is formed, the accretion onto the secondary becomes significant enough to increase the mass ratio.
The value of $\qc$ does not differ dramatically depending on gas temperature as long as $c_s/v_{\rm K}<1$. 

We also estimate the long-term evolution of a seed binary analytically. 
Assuming that equation~(\ref{eq:self_similar}) holds, 
we argue that the evolution of binary system would be self-similar, and the short-term evolution of $q(t)$ from our simulations 
can be reused just by updating the initial mass ratio.
As a result, we find that the binary separation is a monotonically increasing function of time for any $q_0$. 
This result suggests that close binaries are difficult to form.
In the future, we will include direct computations of binary orbit in our simulations in order to investigate the effect of binary growth by accretion.

\section*{Acknowledgments}
This work is partially based on the Master's thesis of Suguru Tanaka \citep{Tanaka_MS}.
We are grateful to Volker Springel for providing us with the original version of {\tt GADGET-3} code,
 and to Kengo Tomida and Fumio Takahara for useful discussions and continuous encouragement.
KN acknowledges the partial support by JSPS KAKENHI Grant Number 26247022.
Numerical simulations were in part carried out on XC30 at the Centre for Computational Astrophysics, National Astronomical Observatory of Japan.

\section*{Appendix}\label{sec:appendix}
\subsection*{Resolution dependence of mass ratio evolution}

\begin{figure}
  \begin{center}
    \includegraphics[width=\columnwidth]{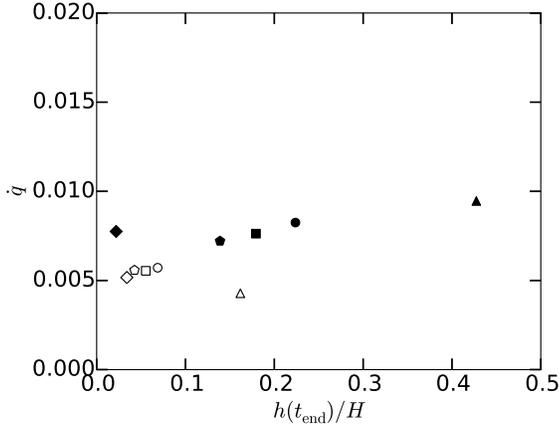}
  \end{center}
  \caption{Mean rate of mass ratio change ${\dot q}$ versus $h(t_{\rm end})/H$ which is the ratio of SPH smoothing length to the scale height at the outer edge of the circum-secondary disc.  The open symbols are for hot case, and the filled symbols are for the cold case
  with $N_{\rm SPH}=64^3$ (triangle), 128$^3$(circle), $2\times 128^3$ (square), $4\times 128^3$ (pentagon), 256$^3$ (diamond). 
  The value of $h(t_{\rm end})/H$ is smaller for better resolution.   All of our simulations presented in the main text is performed with $N_{\rm SPH}=128^3$.}
  \label{fig:hoverH}
\end{figure}

In order to investigate the resolution dependence of mass ratio evolution, we rerun several simulations in the cases with $q_0=0.7$ (hot and cold) and varying number of SPH particles $N_{\rm SPH}=64^3,\,128^3,\,2\times 128^3,\,4\times128^3,$ and 256$^3$.
In a steady circum-stellar disc, pressure gradient and gravitational force are in equilibrium in the vertical direction as mentioned in Subsection~\ref{subsec:results_disc}.
Therefore, the scale height of this disc $H$ is represented by
\begin{equation}
H(r_{\rm cyl}) = \frac{c_s}{v_{\rm K}(r_{\rm cyl})} r_{\rm cyl} .
\end{equation}
We introduce $h/H$, the ratio of SPH smoothing length to the scale height at the outer edge of the circum-secondary disc.
Here, we define the edge of the circum-secondary disc where $J=U_{\rm L1}$.
In Fig.~\ref{fig:hoverH}, we plot the mean rate of change of the mass ratio, ${\dot q}=(q(t_{\rm end})-q_0)/t_{\rm end}$, as a function of $h(t_{\rm end})/H$.  In both hot and cold cases, one can see that the variation of ${\dot q}$ is within $25 \%$ for different resolution.
\cite{Young_etal_15} suggested that ${\dot q}$ is independent of resolution if the SPH smoothing length at the outer edge of circum-secondary disc satisfies $h/H<1$. 
Indeed, we also confirm this criterion with our three dimensional simulations.

\bibliography{ref}

\begin{thebibliography}{}

\bibitem[\protect\citeauthoryear{{Adams}, {Ruden} \& {Shu}}{{Adams}
  et~al.}{1989}]{Adams_etal_89}
{Adams} F.~C.,  {Ruden} S.~P.,    {Shu} F.~H.,  1989, ApJ, 347, 959

\bibitem[\protect\citeauthoryear{{Artymowicz} \& {Lubow}}{{Artymowicz} \&
  {Lubow}}{1996}]{Artymowicz_Lubow_96}
{Artymowicz} P.,  {Lubow} S.~H.,  1996, ApJ, 467, L77

\bibitem[\protect\citeauthoryear{{Attwood}, {Goodwin}, {Stamatellos} \&
  {Whitworth}}{{Attwood} et~al.}{2009}]{Attwood_etal_09}
{Attwood} R.~E.,  {Goodwin} S.~P.,  {Stamatellos} D.,    {Whitworth} A.~P.,
  2009, A\&A, 495, 201

\bibitem[\protect\citeauthoryear{{Bate}}{{Bate}}{1997}]{Bate_97}
{Bate} M.~R.,  1997, MNRAS, 285, 16

\bibitem[\protect\citeauthoryear{{Bate}}{{Bate}}{2000}]{Bate_00}
{Bate} M.~R.,  2000, MNRAS, 314, 33

\bibitem[\protect\citeauthoryear{{Bate}}{{Bate}}{2009a}]{Bate_09a}
{Bate} M.~R.,  2009a, MNRAS, 392, 590

\bibitem[\protect\citeauthoryear{{Bate}}{{Bate}}{2009b}]{Bate_09b}
{Bate} M.~R.,  2009b, MNRAS, 392, 1363

\bibitem[\protect\citeauthoryear{{Bate} \& {Bonnell}}{{Bate} \&
  {Bonnell}}{1997}]{Bate_Bonnell_97}
{Bate} M.~R.,  {Bonnell} I.~A.,  1997, MNRAS, 285, 33

\bibitem[\protect\citeauthoryear{{Bate}, {Bonnell} \& {Bromm}}{{Bate}
  et~al.}{2002a}]{Bate_etal_02b}
{Bate} M.~R.,  {Bonnell} I.~A.,    {Bromm} V.,  2002a, MNRAS, 332, L65

\bibitem[\protect\citeauthoryear{{Bate}, {Bonnell} \& {Bromm}}{{Bate}
  et~al.}{2002b}]{Bate_etal_02c}
{Bate} M.~R.,  {Bonnell} I.~A.,    {Bromm} V.,  2002b, MNRAS, 336, 705

\bibitem[\protect\citeauthoryear{{Bate}, {Bonnell} \& {Bromm}}{{Bate}
  et~al.}{2003}]{Bate_Bonnell_Bromm_03}
{Bate} M.~R.,  {Bonnell} I.~A.,    {Bromm} V.,  2003, MNRAS, 339, 577

\bibitem[\protect\citeauthoryear{{Bonnell}}{{Bonnell}}{1994}]{Bonnell_94}
{Bonnell} I.~A.,  1994, MNRAS, 269

\bibitem[\protect\citeauthoryear{{Bonnell} \& {Bate}}{{Bonnell} \&
  {Bate}}{1994}]{Bonnell_Bate_94b}
{Bonnell} I.~A.,  {Bate} M.~R.,  1994, MNRAS, 271

\bibitem[\protect\citeauthoryear{{Boss} \& {Bodenheimer}}{{Boss} \&
  {Bodenheimer}}{1979}]{Boss_Bodenheimer_79}
{Boss} A.~P.,  {Bodenheimer} P.,  1979, ApJ, 234, 289

\bibitem[\protect\citeauthoryear{{De Rosa}, {Patience}, {Wilson}, {Schneider},
  {Wiktorowicz}, {Vigan}, {Marois}, {Song}, {Macintosh}, {Graham}, {Doyon},
  {Bessell}, {Thomas} \& {Lai}}{{De Rosa} et~al.}{2014}]{DeRosa_etal_14}
{De Rosa} R.~J.,  {Patience} J.,  {Wilson} P.~A.,  {Schneider} A.,
  {Wiktorowicz} S.~J.,  {Vigan} A.,  {Marois} C.,  {Song} I.,  {Macintosh} B.,
  {Graham} J.~R.,  {Doyon} R.,  {Bessell} M.~S.,  {Thomas} S.,    {Lai} O.,
  2014, MNRAS, 437, 1216

\bibitem[\protect\citeauthoryear{{D'Orazio}, {Haiman} \&
  {MacFadyen}}{{D'Orazio} et~al.}{2013}]{D'Orazio_etal_13}
{D'Orazio} D.~J.,  {Haiman} Z.,    {MacFadyen} A.,  2013, MNRAS, 436, 2997

\bibitem[\protect\citeauthoryear{{Dunhill}, {Cuadra} \& {Dougados}}{{Dunhill}
  et~al.}{2015}]{Dunhill_etal_15}
{Dunhill} A.~C.,  {Cuadra} J.,    {Dougados} C.,  2015, MNRAS, 448, 3545

\bibitem[\protect\citeauthoryear{{Duquennoy} \& {Mayor}}{{Duquennoy} \&
  {Mayor}}{1991}]{Duquennoy_Mayor_91}
{Duquennoy} A.,  {Mayor} M.,  1991, A\&A, 248, 485

\bibitem[\protect\citeauthoryear{{Farris}, {Duffell}, {MacFadyen} \&
  {Haiman}}{{Farris} et~al.}{2014}]{Farris_etal_14}
{Farris} B.~D.,  {Duffell} P.,  {MacFadyen} A.~I.,    {Haiman} Z.,  2014, ApJ,
  783, 134

\bibitem[\protect\citeauthoryear{{Ghez}, {Neugebauer} \& {Matthews}}{{Ghez}
  et~al.}{1993}]{Ghez_etal_93}
{Ghez} A.~M.,  {Neugebauer} G.,    {Matthews} K.,  1993, ApJ, 106, 2005

\bibitem[\protect\citeauthoryear{{Hanawa}, {Ochi} \& {Ando}}{{Hanawa}
  et~al.}{2010}]{Hanawa_etal_10}
{Hanawa} T.,  {Ochi} Y.,    {Ando} K.,  2010, ApJ, 708, 485

\bibitem[\protect\citeauthoryear{{Kouwenhoven}, {Brown}, {Zinnecker}, {Kaper}
  \& {Portegies Zwart}}{{Kouwenhoven} et~al.}{2005}]{Kouwenhoven_etal_05}
{Kouwenhoven} M.~B.~N.,  {Brown} A.~G.~A.,  {Zinnecker} H.,  {Kaper} L.,
  {Portegies Zwart} S.~F.,  2005, A\&A, 430, 137

\bibitem[\protect\citeauthoryear{{Kraus}, {Ireland}, {Martinache} \&
  {Hillenbrand}}{{Kraus} et~al.}{2011}]{Kraus_etal_11}
{Kraus} A.~L.,  {Ireland} M.~J.,  {Martinache} F.,    {Hillenbrand} L.~A.,
  2011, ApJ, 731, 8

\bibitem[\protect\citeauthoryear{{Matsumoto}, {Hanawa} \&
  {Nakamura}}{{Matsumoto} et~al.}{1997}]{Matsumoto_etal_97}
{Matsumoto} T.,  {Hanawa} T.,    {Nakamura} F.,  1997, ApJ, 478, 569

\bibitem[\protect\citeauthoryear{{Miyama}, {Hayashi} \& {Narita}}{{Miyama}
  et~al.}{1984}]{Miyama_etal_84}
{Miyama} S.~M.,  {Hayashi} C.,    {Narita} S.,  1984, ApJ, 279, 621

\bibitem[\protect\citeauthoryear{{Narita}, {Hayashi} \& {Miyama}}{{Narita}
  et~al.}{1984}]{Narita_etal_84}
{Narita} S.,  {Hayashi} C.,    {Miyama} S.~M.,  1984, Progress of Theoretical
  Physics, 72, 1118

\bibitem[\protect\citeauthoryear{{Ochi}, {Sugimoto} \& {Hanawa}}{{Ochi}
  et~al.}{2005}]{Ochi_etal_05}
{Ochi} Y.,  {Sugimoto} K.,    {Hanawa} T.,  2005, ApJ, 623, 922

\bibitem[\protect\citeauthoryear{{Offner}, {Klein}, {McKee} \&
  {Krumholz}}{{Offner} et~al.}{2009}]{Offner_etal_09}
{Offner} S.~S.~R.,  {Klein} R.~I.,  {McKee} C.~F.,    {Krumholz} M.~R.,  2009,
  ApJ, 703, 131

\bibitem[\protect\citeauthoryear{{Price}}{{Price}}{2007}]{Price_07}
{Price} D.~J.,  2007, Publ. Astron. Soc. Australia, 24, 159

\bibitem[\protect\citeauthoryear{{Raghavan}, {McAlister}, {Henry}, {Latham},
  {Marcy}, {Mason}, {Gies}, {White} \& {ten Brummelaar}}{{Raghavan}
  et~al.}{2010}]{Raghavan_etal_10}
{Raghavan} D.,  {McAlister} H.~A.,  {Henry} T.~J.,  {Latham} D.~W.,  {Marcy}
  G.~W.,  {Mason} B.~D.,  {Gies} D.~R.,  {White} R.~J.,    {ten Brummelaar}
  T.~A.,  2010, ApJS, 190, 1

\bibitem[\protect\citeauthoryear{{Saigo} \& {Hanawa}}{{Saigo} \&
  {Hanawa}}{1998}]{Saigo_Hanawa_98}
{Saigo} K.,  {Hanawa} T.,  1998, ApJ, 493, 342

\bibitem[\protect\citeauthoryear{{Shakura} \& {Sunyaev}}{{Shakura} \&
  {Sunyaev}}{1973}]{Shakura_Sunyaev_73}
{Shakura} N.~I.,  {Sunyaev} R.~A.,  1973, A\&A, 24, 337

\bibitem[\protect\citeauthoryear{{Springel}}{{Springel}}{2005}]{Springel_05}
{Springel} V.,  2005, MNRAS, 364, 1105

\bibitem[\protect\citeauthoryear{{Tanaka}}{{Tanaka}}{2010}]{Tanaka_MS}
{Tanaka} S.,  2010, Master's thesis, Osaka University

\bibitem[\protect\citeauthoryear{{Tohline}}{{Tohline}}{2002}]{Tohline_02}
{Tohline} J.~E.,  2002, ARA\&A, 40, 349

\bibitem[\protect\citeauthoryear{{Tsuribe} \& {Inutsuka}}{{Tsuribe} \&
  {Inutsuka}}{1999a}]{Tsuribe_Inutsuka_99b}
{Tsuribe} T.,  {Inutsuka} S.,  1999a, ApJ, 526, 307

\bibitem[\protect\citeauthoryear{{Tsuribe} \& {Inutsuka}}{{Tsuribe} \&
  {Inutsuka}}{1999b}]{Tsuribe_Inutsuka_99a}
{Tsuribe} T.,  {Inutsuka} S.,  1999b, ApJ, 523, L155

\bibitem[\protect\citeauthoryear{{Vorobyov}}{{Vorobyov}}{2010}]{Vorobyov_10}
{Vorobyov} E.~I.,  2010, ApJ, 723, 1294

\bibitem[\protect\citeauthoryear{{Williams} \& {Tohline}}{{Williams} \&
  {Tohline}}{1988}]{Williams_Tohline_88}
{Williams} H.~A.,  {Tohline} J.~E.,  1988, ApJ, 334, 449

\bibitem[\protect\citeauthoryear{{Woodward}, {Tohline} \& {Hachisu}}{{Woodward}
  et~al.}{1994}]{Woodward_etal_94}
{Woodward} J.~W.,  {Tohline} J.~E.,    {Hachisu} I.,  1994, ApJ, 420, 247

\bibitem[\protect\citeauthoryear{{Young}, {Baird} \& {Clarke}}{{Young}
  et~al.}{2015}]{Young_etal_15}
{Young} M.~D.,  {Baird} J.~T.,    {Clarke} C.~J.,  2015, MNRAS, 447, 2907

\bibitem[\protect\citeauthoryear{{Young} \& {Clarke}}{{Young} \&
  {Clarke}}{2015}]{Young_Clarke_15}
{Young} M.~D.,  {Clarke} C.~J.,  2015, MNRAS, 452, 3085

\end{thebibliography}

\end{document}